\documentclass[twocolumn,english,aps,prb,twocolum,superscriptaddress,bibnotes,amsmath,amssymb,floatfix]{revtex4-2}
\usepackage[colorlinks=true,citecolor=blue,linkcolor=magenta]{hyperref}

\usepackage[markup=nocolor, authormarkupposition=left]{changes} 
\usepackage{soul}
\usepackage[utf8]{inputenc}
\usepackage[english]{babel}
\usepackage{amsmath,amsfonts,amssymb}
\usepackage[T1]{fontenc}
\usepackage{url}
\usepackage{amsmath}
\usepackage{amsfonts}
\usepackage{amssymb}
\usepackage{epstopdf}
\usepackage{graphicx}

\begin{document}
\title{Three-dimensional integration enables ultra-low-noise, isolator-free Si photonics}

\author{
Chao Xiang$^{1,2,\ast,\dagger}$,
Warren Jin$^{1,3,\ast}$,
Osama Terra$^{1,\ast}$,
Bozhang Dong$^{1,\ast}$,
Heming Wang$^{1}$,\\
Lue Wu$^4$,
Joel Guo$^1$,
Theodore J. Morin$^1$,
Eamonn Hughes$^5$,
Jonathan Peters$^1$,
Qing-Xin Ji$^4$,
Avi Feshali$^3$,
Mario Paniccia$^3$,
Kerry J. Vahala$^4$,
and John E. Bowers$^{1,5,\dagger}$\\
\textit{$^1$Department of Electrical and Computer Engineering, University of California, Santa Barbara, CA, USA\\
$^2$Department of Electrical and Electronic Engineering, The University of Hong Kong, Hong Kong, China\\
$^3$Anello Photonics, Santa Clara, CA, USA\\
$^4$J. Watson Laboratory of Applied Physics, California Institute of Technology, Pasadena, CA, USA\\
$^5$Materials Department, University of California, Santa Barbara, CA, USA}\\
$^*$These authors contributed equally\\
$^\dagger$Email: cxiang@eee.hku.hk, bowers@ece.ucsb.edu\\
}

\begin{abstract}

While photonic integrated circuits (PICs) are being widely used in applications such as telecommunications and datacenter interconnects, PICs capable of replacing bulk optics and fibers in high-precision, highly-coherent applications will require ultra-low-noise laser sources to be integrated with other photonic components in a compact and robustly aligned format -- that is, on a single chip~\cite{margalit2021perspective,doerr2015silicon,Thomson:16,Cheng:18,Soref:06}. 
Such PICs could offer superior scalability for complex functionalities and volume production, as well as improved stability and reliability over time. However, there are two major issues preventing the realization of such envisioned PICs: the high phase noise of semiconductor lasers, and the difficulty of integrating optical isolators directly on chip. 
PICs are still considered as inferior solutions in optical systems such as microwave synthesizers~\cite{Marpaung:19}, optical gyroscopes~\cite{lefevre2014fiber} and atomic clocks~\cite{ludlow2015optical}, despite their advantages in size, weight, power consumption and cost (SWaPC).
Here, we challenge this convention by introducing three-dimensional (3D) integration in silicon photonics that results in ultra-low-noise, isolator-free PICs.
Through multiple monolithic and heterogeneous processing sequences, direct on-chip integration of III-V gain and ultra-low-loss (ULL) silicon nitride (SiN) waveguides with optical loss around 0.5 dB/m are demonstrated.
Consequently, the demonstrated PIC enters a new regime, such that an integrated ultra-high-Q cavity reduces the laser noise close to that of fiber lasers. 
Moreover, the cavity acts as an effective block for any downstream on-chip or off-chip reflection-induced destabilization, thus eliminating the need for optical isolators. 
We further showcase isolator-free, widely-tunable, low-noise, heterodyne microwave generation using two ultra-low-noise lasers on the same silicon chip.
The 3D integration of feedback-insensitive, ultra-low-noise lasers on ULL PICs marks a critical step towards complex systems and networks on silicon.

\end{abstract}
\maketitle

Following the path of electronic integrated circuits (EICs), silicon photonics promises to enable photonic integrated circuits (PICs) with high densities, advanced functionality, and portability. 
Although various silicon photonics foundries are rapidly developing PIC capabilities -- enabling volume production of modulators, photodetectors, and most recently lasers - Si PICs have yet to achieve the stringent requirements on laser noise and overall system stability imposed by many applications such as microwave oscillators, atomic physics and precision metrology~\cite{siew2021review,Giewont2019300mm,rahim2019open}. Semiconductor lasers must strongly suppress amplified spontaneous emission (ASE) noise to achieve narrow linewidth for these applications~\cite{coldren_diode_2012,henry1982theory}. They will also require isolation from the rest of the optical system, otherwise the laser source will be sensitive to back-reflections from downstream optical components that are beyond the control of the PIC designer~\cite{Tkach1986regimes}. 
In many integrated photonic solutions, a bulk optical isolator must be inserted between the laser chip and the rest of the system, significantly increasing the complexity, as well as the cost of assembly and packaging~\cite{Lee2016photonic}. 

To enrich the capabilities of Si PICs and avoid multi-chip optical packaging, non-group-IV materials need to be heterogeneously integrated to enable crucial devices, including high-performance lasers, amplifiers and isolators~\cite{xiang2021JSTQE,liang2021recent,marshall2018heterogeneous}. It has now been widely acknowledged that III-V materials are required to provide efficient optical gain for semiconductor lasers and amplifiers in silicon photonics regardless of the integration architecture, but concerns still remain for a CMOS fab to incorporate magnetic materials, which are currently used in industry-standard optical isolators~\cite{pintus2017microring}.

Fortunately, a synergistic path towards ultra-low laser noise and low feedback sensitivity exists - using ultra-high-Q cavities for lasers that not only reduce the phase noise, but also enhance the feedback tolerance to downstream links~\cite{santis2014high}. These effects scale with the cavity Q-factor and ultra-high-Q cavities would thus endow integrated lasers with unprecedented coherence and stability \cite{Harfouche:20}. 
The significance is two-fold. First, the direct integration of ultra-low-noise lasers on Si PICs without the need for optical isolators simplifies PIC fabrication and packaging. Furthermore, this approach does not introduce magnetic materials to a CMOS fab since isolators are not obligatory for such complete PICs. 

%%%%%%%%%%%%%%%%%%%%%%%%%%%%%%%%%%%%%%%%%%%%%%%%%%%%%%%%%%%%%%%%%%%%%%
\begin{figure*}[t!]
\centering
\includegraphics{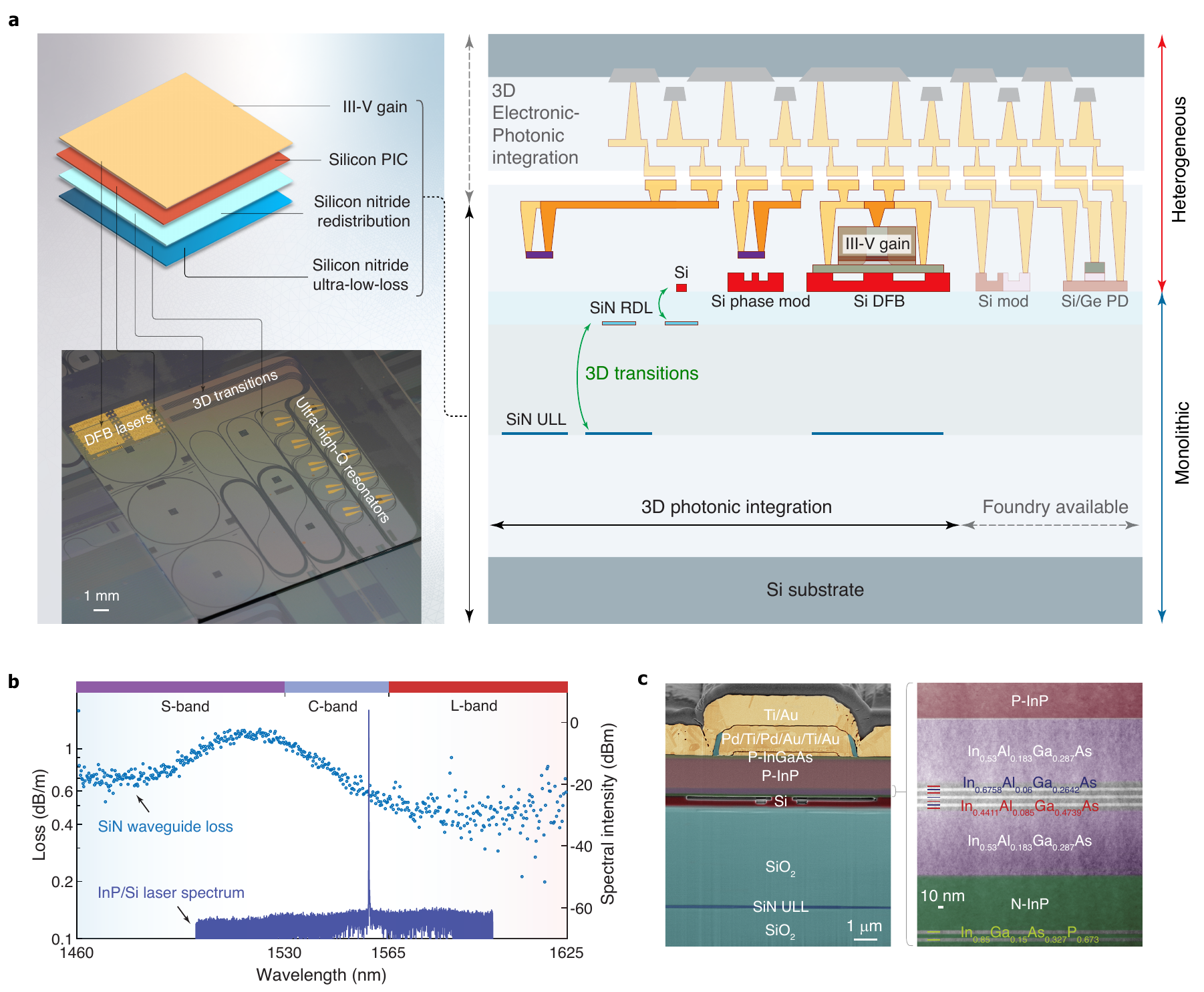}
\caption{\textbf{3D integrated silicon photonic integrated circuit chip.}
\textbf{a}. Concept of 3D photonic integration of functional layers (upper left) and the corresponding devices on a fabricated 3D PIC (device picture shown in the lower left). This chip is singulated from a fully-fabricated 100-mm-diameter wafer. The SiN wafer process is performed on a 200-mm-diameter wafer which was cored into 100-mm-diameter wafers for the heterogeneous laser fabrication. The right plot shows the cross-section of the demonstrated 3D PIC in solid colors.  We envision future works will enable additional functionality, such as integration with foundry-available Si modulators and Ge/Si PDs, and 3D electronic-photonic heterogeneous integration, which are shown in transparent colors. Both monolithic and heterogeneous integration processes are employed, in which 3D transitions are critical to the vertical integration of functionality layers.
\textbf{b}. Measured III-V/Si DFB laser spectrum centered at telecom C-band on the 3D PIC (right axis) and measured ULL SiN waveguide loss (left axis) across the telecom S-,C-, and L-bands on the same 3D PIC.
\textbf{c}. False-colored focused ion beam scanning electron microscopy (FIB-SEM) image of the fabricated 3D PIC showing the laser cross-section (left) and transmission electron microscopy (TEM) image showing the layered InP epitaxy stack after bonding and substrate removal (right).
}
\label{Fig:1}
\end{figure*}
%%%%%%%%%%%%%%%%%%%%%%%%%%%%%%%%%%%%%%%%%%%%%%%%%%%%%%%%%%%%%%%%%%%%%%

\medskip
\noindent \textbf{3D photonic integration of lasers and ULL waveguides}

Now we consider developing an integration architecture and process flow to seamlessly integrate the III-V-based lasers with ULL waveguides. Among various ULL integrated photonics platforms, SiN has emerged as the leading performer and enabled a series of breakthroughs in metrology, sensing and telecommunications~\cite{Kippenberg:18,Gaeta:19,Xiang2022silicon,Dai2012passive}.
To achieve ultra-low waveguide loss, SiN waveguides require high-temperature annealing \cite{Bauters:11,liu2021high,ji2021methods,Puckett2021422} that violates the thermal budget of back-end-of-line (BEOL) semiconductor manufacturing processes.
Front-end-of-line (FEOL) fabricated ULL SiN waveguides are nonetheless susceptible to subsequent processing steps that could introduce additional loss, especially during heterogeneous laser integration, which involves multiple etches and depositions. 
To address these issues, we propose to use 3D structures for the integration of lasers with ULL waveguides. 
Recent years have witnessed the developments of 3D integration in electronics by heterogeneously or monolithically integrating layers for increased circuit densities and functionalities~\cite{shulaker2017three,Rachmady2019300mm}. Unlike electronic ICs that rely on the interlayer metallic vias for interconnects, our 3D Si PIC leverages evanescent coupling across multiple layers and uses waveguide geometry designs to achieve interlayer transitions that are otherwise forbidden. Monolithic 3D photonic integration has been investigated for passive devices~\cite{sacher2018monolithically}, and the heterogeneous integration with l asers for optimized co-designs of actives and passives will fully unlock the potential in enabling complex and high-performance photonic devices and integrated circuits.

We thus effectively separate a heterogeneous Si PIC into layers with respective photonic functionalities as shown in Fig.~\ref{Fig:1}a.
The designed device consists of four major functionality layers, including the III-V gain layer, Si PIC layer, SiN redistribution layer (RDL) and SiN ULL layer.
Such design necessitates inter-layer transition across multiple functional layers. 
More specifically, the separation of Si and ULL SiN layers is approximately 4 \textmu m, such that the ULL SiN layer can be effectively isolated from subsequent Si and InP processing steps, thereby retaining the performance of the ULL SiN (see Supplementary Information). We then introduce a photonic RDL between the Si and ULL SiN layers for the control of coupling and decoupling between the top active layers and the bottom passive layer. As a result, the coupling between the active layers and the SiN ULL layer is determined by the RDL layer. Highly-efficient active-passive layer transition can be provided by the RDL where necessary.  When no mode coupling is provided by RDL, the active layers and passive layers function independently. Furthermore, vertical decoupling can also improve the efficiency of waveguide crossings. This 3D architecture could thus leverage vertical space to relax limitations on device density on the PIC, especially compared to conventional heterogeneous laser integration that still limits the PIC to planar circuit layouts~\cite{Komljenovic:16}.     

The cross-section of the 3D PICs is illustrated in Fig.~\ref{Fig:1}a showing its compatibility with foundry-available Si photonic components including Si modulators, and Si/Ge photodetectors. Additionally, such PICs could be further heterogeneously integrated with EICs for high-density 3D E-PICs. 
In our 3D photonic integration structure, the thick oxide separation forms an effective barrier for back-end loss origins, and ultra-high-Q resonators (with intrinsic Q $\sim$ 50 million at the laser wavelength) are integrated with high-performance III-V/Si distributed feedback (DFB) lasers (Fig.~\ref{Fig:1}b).
It must be noted that the 3D integration can result in multiple overlapping but decoupled photonic functionality layers - a goal not possible in previous demonstrations of heterogeneous integration~\cite{Xiang:20}. This decoupling is now enabled by the large vertical mode separation, which is bridged by the SiN RDL. 
The multilayer structure of the fabricated device and InP epi stack are shown in Fig.~\ref{Fig:1}c.

%%%%%%%%%%%%%%%%%%%%%%%%%%%%%%%%%%%%%%%%%%%%%%%%%%%%%%%%%%%%%%%%%%%%%%
\begin{figure*}[t!]
\centering
\includegraphics[width=6.5in]{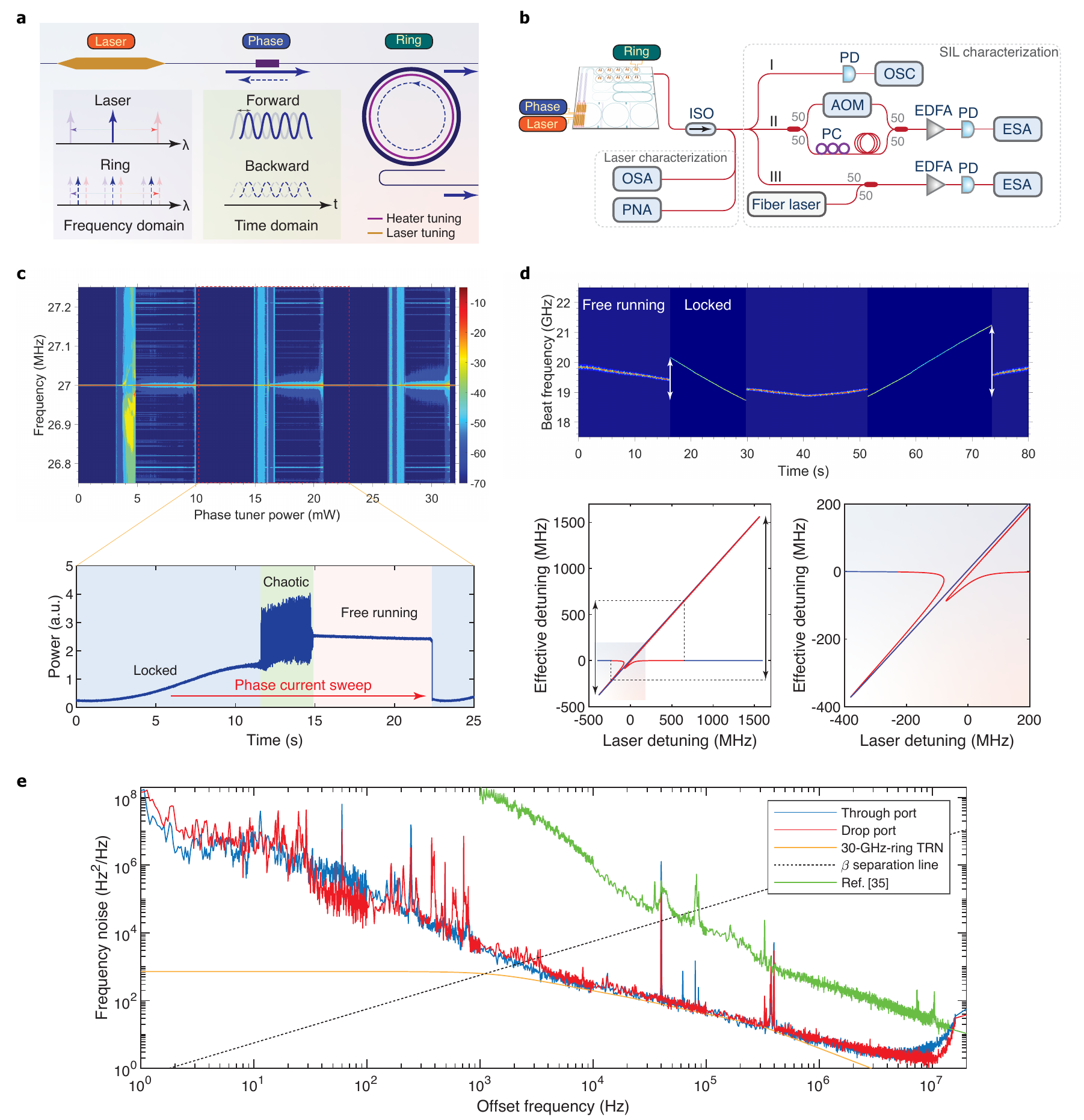}
\caption{\textbf{Laser self-injection locking and phase noise.}
\textbf{a}. Schematic illustration of the laser self-injection locking, which requires tuning in wavelength and phase to work. There are three knobs used to control the working regimes respectively: laser current, phase heater current, and ring heater current.
\textbf{b}. The experimental setup to characterize the laser performance and the self-injection locking process.
OSA: optical spectrum analyzer.
PNA: phase noise analyzer.
ISO: isolator.
AOM: acoustic-optic modulator.
OSC: oscilloscope.
PD: photodetector.
ESA: electrical spectrum analyzer.
EDFA: erbium-doped fiber amplifier.
PC: polarization controller.
\textbf{c}. The dependence of laser self-injection locking on the phase tuner power. The top figure shows the change in delayed self-heterodyne beat spectrum recorded by an ESA and the bottom plot shows the corresponding power recorded on the oscilloscope of one phase tuning period revealing the locked, chaotic, and unlocked states. The AOM used in this experiment has a center frequency of 27 MHz.
\textbf{d}. The laser beat frequency with a fiber laser during the ring resonance blue-shift sweep and red-shift sweep. The vertical arrow marks the self-injection locking range. The lower plot is a calculation of asymmetric laser frequency locking range behavior without thermal crosstalk for the bi-directional sweep. The Blue and red sections of the curve indicate stable and unstable branches, respectively.
\textbf{e}. The frequency noise of the laser output taken from the through port and drop port of the 30-GHz ring resonator. Comparisons also show the thermo-refractive noise trace (TRN) of the 30-GHz-FSR ring resonator and beta separation line. The green curve shows the frequency noise of the SIL laser reported in reference~\cite{xiang2021laser}.
}
\label{Fig:2}
\end{figure*}
%%%%%%%%%%%%%%%%%%%%%%%%%%%%%%%%%%%%%%%%%%%%%%%%%%%%%%%%%%%%%%%%%%%%%%

\medskip
\noindent \textbf{Single-chip self-injection locked ultra-low-noise lasers}

We leverage self-injection locking of InP/Si DFB lasers to thermally-tunable SiN ultra-high-Q resonators for ultra-low-noise lasers on the 3D Si PIC. The working principle of such a device is summarized in Fig.~\ref{Fig:2}a, which requires the laser and ring resonance wavelengths to match in the frequency domain, as well as the forward and backward signals to phase match in the time domain. To set the device to the proper working conditions, the InP/Si laser wavelength is tuned by the applied gain current; the SiN ring resonance is tuned by the thermal heater, and the forward/backward phase is tuned by the thermal phase tuner placed on the Si waveguides. Once both wavelength matching and phase matching are achieved, the free-running laser locks to the ultra-high-Q resonator thanks to Rayleigh back-scattering, resulting in several resonator-defined laser properties.

We investigate the dynamics and performance of the self-injection locked (SIL) laser using the measurement setup as shown in Fig.~\ref{Fig:2}b. 
Due to the availability of an on-chip phase tuner between the laser and ring resonator, we can clearly unveil the phase-dependent locking dynamics. In previous butt-coupled SIL experiments, tuning the chip-to-chip phase also varies the coupling loss, i.e. the output power. Because the InP/Si laser and SiN resonator are heterogeneously integrated together and the phase is thermally tuned on chip, these are now decoupled in our experiment. 
Figure \ref{Fig:2}c plots the dependence of laser coherence on the phase tuner power causing the phase shift. The laser wavelength is preset to match one of the ring resonances. We can observe a periodic dependence of the laser coherence when the laser-to-resonator phase is tuned by several one-direction $\pi$ periods. Within each period, the laser goes through the low-phase-noise locking, chaotic coherence collapse, and high-phase-noise free running regimes. These regimes are also observed from the time-domain power trace recorded on an oscilloscope when the current on the phase tuner is swept across a full period (Fig.~\ref{Fig:2}c bottom).

The ring resonance is another degree of freedom to control the locking dynamics. By tuning the thermal tuner current on the ULL SiN ring in both directions, the laser can be switched between the free-running state and the locked state as plotted in Fig.~\ref{Fig:2}d top. Depending on the phase, the locking range can be different for the bi-directional sweep. We observed about 1.4 GHz and 2.4 GHz locking ranges for the bi-directional sweep. This measured locking range is also affected by the thermal crosstalk during the resonance tuning, as evidenced by the laser frequency shift at free running state. Fig.~\ref{Fig:2}d bottom plots the modeled asymmetric locking range without the thermal crosstalk at phase condition. Details of the calculation can be found in Supplementary Information. 

Ultra-low laser frequency noise enabled by self-injection locking has been extensively studied in recent years~\cite{kondratiev2022recent}. These demonstrations, however, mostly rely on individual ultra-high-Q resonators, including crystalline WGM resonators \cite{Liang:15b} and SiN ring \cite{Jin2021hertz} or spiral resonators \cite{li2021reaching}. The laser and the resonator are thus separate and need free-space or fiber coupling. We recently demonstrated SIL of lasers with dispersion-engineered resonators on a heterogeneous chip for soliton microcomb generation~\cite{xiang2021laser}, but the laser frequency noise is still quite high, especially in the range of 1 kHz to 100 kHz which are critical to microwave and sensing applications~\cite{leeson1966short}. Our current device, with around 0.5 dB/m ultra-low loss integrated with lasers on the same chip, demonstrated the lowest laser frequency noise for a single-chip device, with around 250 Hz$^2$/Hz and 2.3 Hz$^2$/Hz at 10 kHz offset and at the white noise floor, respectively, for the through port. The white noise floor for the drop port is even lower (1.7 Hz$^2$/Hz), exhibiting about 5 Hz fundamental linewidth. It needs to be noted that these results are achieved with a relatively compact 30-GHz-FSR resonator, and the laser FN is limited by the thermorefractive noise (TRN). By using a larger ring radius or a spiral-shaped resonator to reduce the TRN, lower FN (e.g. sub-Hz fundamental laser linewidth) should be achieved using the same design strategy and fabrication process. 
%%%%%%%%%%%%%%%%%%%%%%%%%%%%%%%%%%%%%%%%%%%%%%%%%%%%%%%%%%%%%%%%%%%%%%
\begin{figure*}[t!]
\centering
\includegraphics[width=6.5in]{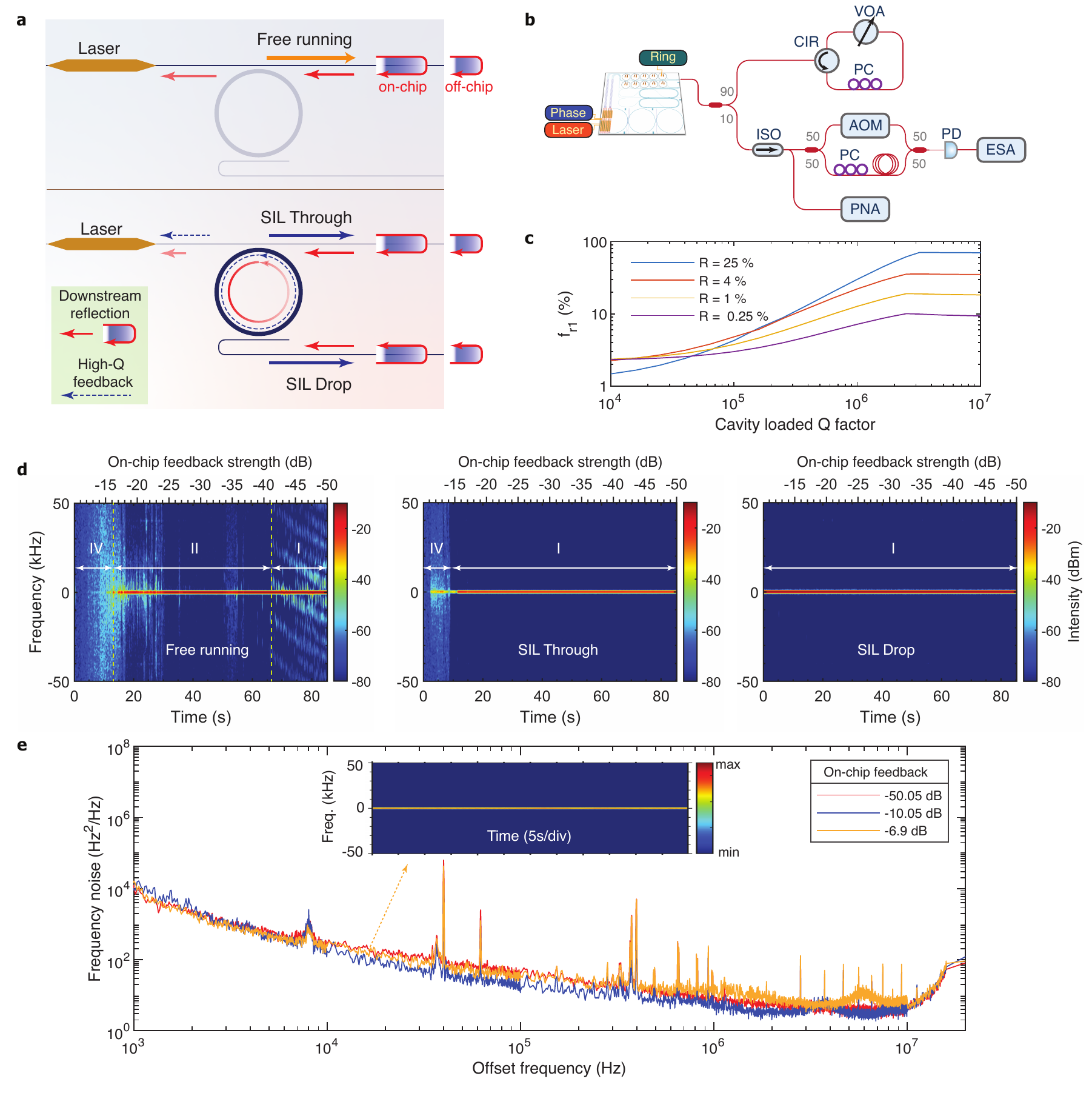}
\caption{\textbf{Feedback insensitivity of the self-injection locked laser.}
\textbf{a}. Schematic illustration of the feedback influence for the laser working at free running state and self-injection locking state. Under self-injection locking conditions, both through port and drop port are characterized.
\textbf{b}. Experimental setup for the feedback sensitivity characterization. 
\textbf{c}. Calculation of the dependence of critical feedback level (the highest tolerable reflection) for Regime I boundary ($f_{r1}$) on the cavity-loaded device. The backscatter (R) from the UHQ resonator also impacts the highest tolerable downstream reflections.
\textbf{d}. The laser spectral lineshape evolution recorded with an ESA by self-heterodyning with an AOM for the free-running laser state (left), SIL through port output (middle), and SIL drop port output (right). Different feedback regimes are indicated and details of the regimes are covered in the Supplementary Information.
\textbf{e}. Frequency noise of the drop port at SIL condition under different on-chip feedback levels. The inset shows the recorded laser spectral lineshape evolution under a maximum -6.9 dB on-chip feedback.
}
\label{Fig:3}
\end{figure*}
%%%%%%%%%%%%%%%%%%%%%%%%%%%%%%%%%%%%%%%%%%%%%%%%%%%%%%%%%%%%%%%%%%%%%%

\medskip
\noindent \textbf{Cavity-mediated feedback sensitivity and isolator-free operation}

In addition to frequency noise, integration with the ultra-high-Q cavity dramatically reduces the feedback sensitivity~\cite{schunk1988numerical}. This goal has been pursued by many demonstrations, but due to the difficulty of integrating ultra-high-Q cavities with lasers, the feedback tolerance is limited such that an isolator is still required to operate in the strong feedback regime (> -10 dB)~\cite{zhang2020high,gomez2020high}. Here, thanks to the attained ultra-high-Q device that decouples efficiently the intra-cavity field and the reflected field, the locked DFB laser thus gains significant feedback tolerance.

In the current SIL configuration with an add-drop ring resonator, the laser output can be taken from both through and drop ports (Fig.~\ref{Fig:3}a). The ring resonator itself acts as an intensity filter for both forward output and backward reflections. This results in another degree of freedom in controlling the feedback sensitivity by modifying the loading factor of the ring resonator. The dependence on the feedback is characterized using the experimental setup shown in Fig.~\ref{Fig:3}b. The downstream feedback results in the change of laser coherence for a feedback-sensitive laser. The laser can operate in several different regimes depending on the feedback strength~\cite{Tkach1986regimes}. Stable operation requires the laser to stay in regime I where the laser coherence is maintained. With an increased feedback level, the laser transitions to Regime II, where the linewidth is governed by the feedback phase (the length of the external cavity). The critical feedback level at the boundary of Regime I and II ($f_{r1}$) represents the highest feedback level a laser can tolerate to maintain stable operation. After the laser enters Regime IV, the laser coherence collapses. Our laser did not enter Regime III, in which a significant frequency stabilization due to external optical feedback can take place, regardless of the feedback phase. In general, Regime III is too narrow to be observed in most semiconductor lasers. 

We calculated the critical feedback level as a function of the cavity loaded Q-factor (Fig.~\ref{Fig:3}c). Subject to different Rayleigh backscattering strengths (R), the laser undergoes variable tolerance to downstream reflection. Generally, large high-Q feedback (the Rayleigh scattering from the UHQ resonator) is beneficial in leading to high downstream reflection tolerance. This effect saturates at certain loaded Q-factors when the phase response provided by the resonator cannot compensate for larger reflected powers outside the resonator. 

To experimentally verify the high feedback tolerance due to the integrated laser and ultra-high-Q resonator, we studied the laser coherence dynamics with varied downstream reflection strength, and the results are summarized in Fig.~\ref{Fig:3}d. For the free-running state, the laser enters Regime II at an on-chip feedback level of -41 dB. This level of feedback can occur in typical waveguide couplers and splitters. As a result, such feedback sensitivity puts a stringent requirement on the on-chip or off-chip device design if isolators are removed. On the contrary, SIL with a high-Q cavity at both through and drop ports sees a clear extended Regime I. The critical feedback level for the Regime I boundary is increased to -15 dB and > -10 dB respectively. We further increased the downstream on-chip feedback level to the SIL drop port to -6.9 dB (limited by the chip-to-fiber coupling loss) and observed a stable and constant laser linewidth - the same as obtained below -50 dB downstream reflections (Fig.~\ref{Fig:3}e inset). Such 26 dB and over 34 dB improvement in the feedback sensitivity can enable direct on-chip laser integration with downstream devices that introduce strong reflection, such as Bragg grating filters, FP resonant cavities, and end-fire coupled components, significantly enriching the complexity of fully on-chip optical systems~\cite{guo2022chip,McLemore2022Miniaturizing,Jin2022Micro}. The frequency noise under different feedback strengths from -50 dB to -6.9 dB is also summarized in Fig.~\ref{Fig:3}e. 

\medskip
\noindent \textbf{Widely-tunable heterodyne microwave frequency generation}

%%%%%%%%%%%%%%%%%%%%%%%%%%%%%%%%%%%%%%%%%%%%%%%%%%%%%%%%%%%%%%%%%%%%%%
\begin{figure*}[t!]
\centering
\includegraphics{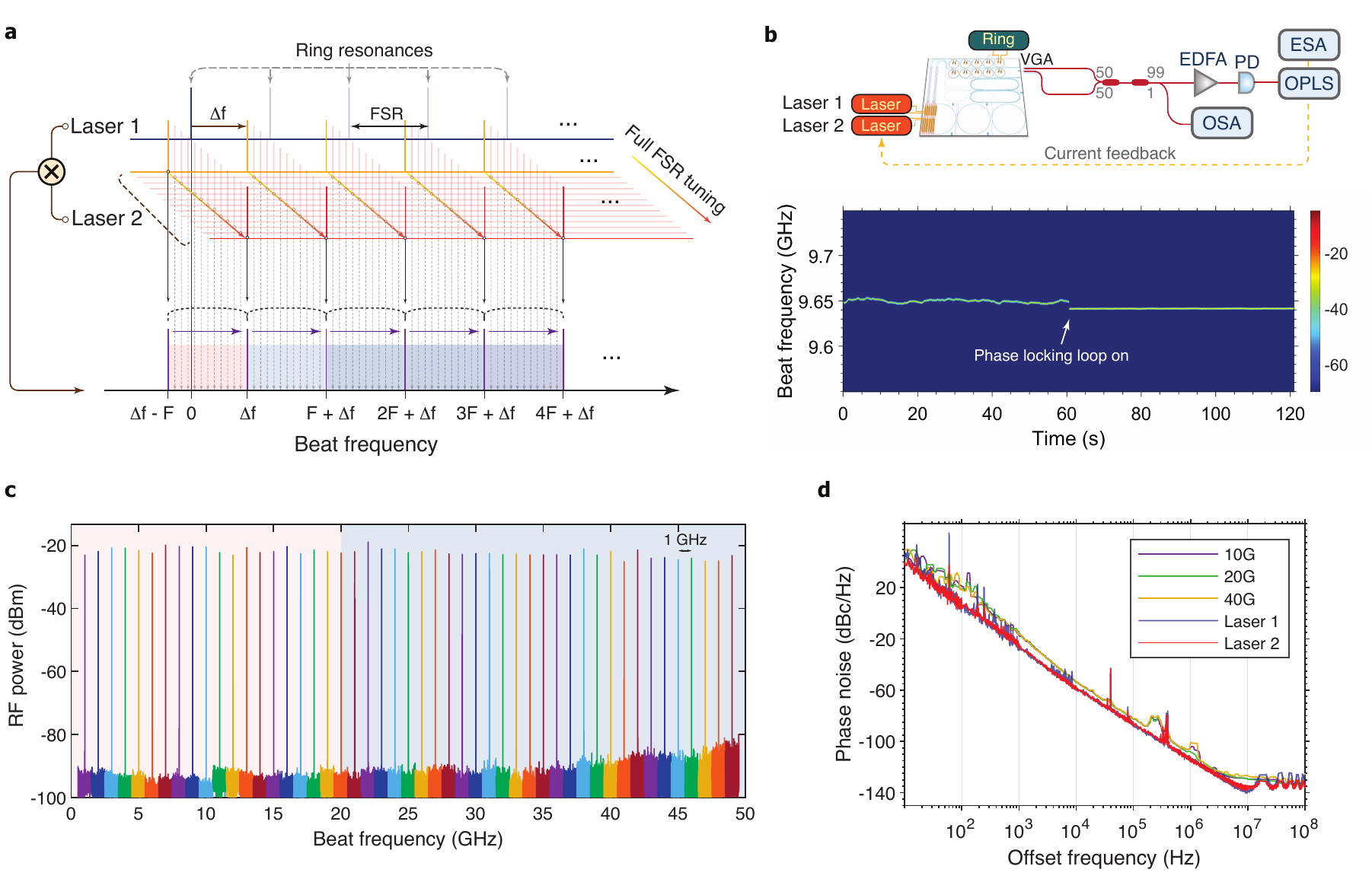}
\caption{\textbf{Widely-tunable microwave signal generation.}
\textbf{a}. Working principle illustration of the widely-tunable heterodyne microwave generation based on two SIL lasers. F denotes FSR and $\Delta f$  is the resonance offset of the two ring resonators without tuning. Shaded-color regions indicate generated microwave signal frequency ranges for a full FSR tuning.
\textbf{b}. Experimental setup for the isolator-free, widely-tunable heterodyne microwave generation. 
OPLS: offset phase lock servo.
VGA: fiber V-groove array.
\textbf{c}. The long-term stability improvement of microwave frequency generation using phase locking loop.
\textbf{d}. Microwave frequency generated by tuning one SIL laser while keeping the other SIL laser at a fixed wavelength. The highest generated microwave frequency is limited by the PD used in this experiment.
\textbf{e}. Carrier-frequency-independent phase noise of the generated microwave signal. The heterodyne microwave signal phase noise is directly determined by the two beating SIL lasers, regardless of the laser frequency separation, i.e. the generated microwave frequency.
}
\label{Fig:4}
\end{figure*}
%%%%%%%%%%%%%%%%%%%%%%%%%%%%%%%%%%%%%%%%%%%%%%%%%%%%%%%%%%%%%%%%%%%%%%

The capability of integrating ultra-low-noise lasers at the wafer scale opens up the possibility of enabling photonic devices that were impractical to integrate. For instance, microwave frequency can be generated by the heterodyne beating of two low-noise lasers on a photodetector (PD) with laser frequency offset at the microwave range~\cite{Hulme2017fully,kittlaus2021low}. The generated frequency could be easily tuned by tuning the laser frequency. This schematic is illustrated in Fig.~\ref{Fig:4}a. Historically, strong semiconductor laser noise prohibited low-noise microwave frequency synthesis using this scheme. Our demonstrated ultra-low-noise lasers provide a route for heterodyne microwave frequency synthesis on a fast PD directly on a PIC, without additional off-chip linewidth narrowing. The advantage of feedback insensitivity is also critical in direct on-chip microwave synthesis since several components including the 3-dB couplers and photodiodes need to follow the lasers and are potentially strong sources of on-chip reflection. To verify the feasibility of our lasers for heterodyne microwave synthesis, we performed a tunable microwave synthesis experiment as shown in Fig.~\ref{Fig:4}b. An optical phase-locked loop on driving the laser current can be used to improve long-term stability, as shown in the bottom inset. This stability could be further improved by the chip packaging. The microwave frequency tuning is achieved by tuning the ring resonances of one ring resonator while keeping the other ring resonance fixed. After laser SIL, the generated microwave signal frequency is determined by the frequency offset of the two resonances. The generated tunable frequency range is ultimately limited by the PD bandwidth since multiple ring resonances separated by ring FSRs can be used for the locking. The laser wavelength tuning range by the injected gain current is a few nanometers of a typical DFB laser depending on the laser cavity length. For the current lasers, we achieved > 3 nm wavelength separation for the two SIL lasers, corresponding to > 375 GHz heterodyne frequency (see Supplementary Information). The microwave signal intensity, while affected by the responsivity of the fast PD and the coupling loss in the current off-chip characterization, could be improved by using directly on-chip III-V amplifiers and waveguides/splitters that are fully compatible with our 3D PIC~\cite{davenport2016heterogeneous}.

The generated microwave signals with frequency tuning from 0-50 GHz at 1-GHz spacing are summarized in Fig.~\ref{Fig:4}c. The frequency tuning is continuous and determined by the thermal phase tuner control on the ring resonator. Using a PD with higher bandwidth, we can further extend the generated frequency tuning range. We characterized the phase noise of the generated microwave signal at different frequencies as shown in Fig.~\ref{Fig:4}d. It clearly shows that the microwave signal phase noise is determined by the laser phase noise and is invariant across the microwave carrier frequencies. This unique advantage of heterodyne microwave signal synthesis is especially suitable for widely-tunable frequency synthesis without noise penalty at high frequencies and provides a practical route for low-noise millimeter-wave and terahertz (THz) generation. Furthermore, no isolators are used in the experiment, which shows the feedback insensitivity could greatly simplify the system architecture and permit a fully on-chip integrated microwave synthesizer when couplers and PDs are integrated on the same 3D Si PIC~\cite{Campbell2022evolution}.

The microwave generation prototype can be optimized using the same integration platform. For example, locking lasers to the same resonator can take advantage of common noise rejection~\cite{Li:14}. 
Such schemes can be explored to gain additional several orders of magnitude reduction in phase noise in our heterodyne microwave synthesizer.

\medskip
\noindent \textbf{Discussion and outlook}

The demonstrated 3D integration of lasers and ultra-low-loss waveguides leverages the advantages of evanescent coupling for vertically-spaced photonic functionality layers. This architecture provides a new design space for complex on-chip photonic systems without being constrained by in-plane process incompatibility and performance degradation. Many optical devices and systems based on the integration of lasers with optical fibers or separate chips nowadays can be translated onto a Si chip using our demonstrated 3D laser integration with ULL technologies, including Brillouin lasers~\cite{Gundavarapu:19}, erbium-doped amplifiers~\cite{liu2022a}, optical gyroscopes~\cite{Liang:17} and optical frequency synthesizers~\cite{Spencer:18}. Currently, the adiabatic taper for 3D transition between the RDL and ULL SiN is made sufficiently long for complete mode transfer between the two SiN layers. Indeed, novel coupling strategies for the waveguides and ultra-high-Q resonators could be explored to further reduce the PIC footprint, including direct waveguide-resonator evanescent tap coupling. Moreover, 3D integration could break the mismatch in the device footprint between different waveguide platforms and use the vertical space to improve the device scalability. For example, ULL waveguides with large bending radius will generally occupy a large device footprint if integrated with compact high-density in-plane Si circuits. With 3D integration, ULL waveguides would not waste mask space, and the saved space could be critical in improving the III-V and Si circuit complexity and scalability. 

The addition of feedback-insensitive ultra-low-noise lasers to silicon photonics will expand the volume production of silicon photonics foundries into applications that remain at small scales. 
Additional optical components such as micro-mirrors could be directly end-fire coupled on our demonstrated isolator-free PIC to enable fully-integrated lasers with Hz-level integrated linewidth~\cite{guo2022chip}. 
As 3D integration provides an effective solution for multiple functionality layers without compromising performance, more materials and functionalities can be added to the existing integrated platform following certain fabrication process guidelines. These materials include lithium niobate~\cite{Zhu:21}, silicon carbide~\cite{Yi2022SiC}, aluminum nitride~\cite{Xiong:12}, III-V quantum-dot materials ~\cite{shang2021perspectives} and so on. For ultra-wideband applications that work across the visible to infrared, 3D integration also provides the potential for materials to be used based on the functionalities instead of transparency limitations since the multi-band applications can be performed at different functionality layers on the same chip. Our demonstration fuels such explorations and enables new building blocks in integrated photonics. Furthermore, 3D heterogeneous integration with electronics can unite the developments of 3D EICs to enable a 3D E-PIC ecosystem and lay the foundation of a whole new class of Si chips. 

\medskip
\begin{footnotesize}
%\begin{methods}

% \noindent \textbf{Funding Information}: 

\noindent \textbf{Acknowledgments}: 
This work is supported by the Defense Advanced Research Projects Agency (DARPA) MTO GRYPHON and LUMOS programs. We thank Andy Netherton, Mingxiao Li, Franklyn Quinlan and Gordon Keeler for their helpful discussions. O.T. acknowledges support from the Fulbright Scholar Program.

\noindent \textbf{Author contribution}: 
C.X. led the 3D PIC device design, fabrication, and characterization. 
W.J. designed the SiN devices and 3D couplers. W.J., A.F. and M.P. handled the SiN wafer processing.
C.X. and J.P. fabricated the 3D PIC device with assistance from W.J..
O.T. and C.X. characterized and gathered the experimental data from the device, including laser noise, locking ranges, phase tuning and microwave generation, with contributions from J.G., B.D., T.M. and Q.J..
B.D., O.T., and C.X. performed the feedback sensitivity measurement.
H.W. provided theoretical calculations and analysis on the locking dynamics and feedback sensitivity.
L.W. performed SIMS concentration analysis of the device.
E.H. took the FIB-SEM and TEM images of the device.
C.X. wrote the manuscript with inputs from W.J., H.M., B.D., O.T. and J.G.. 
All authors commented on and edited the manuscript.
K.J.V and J.E.B supervised the project.

\noindent \textbf{Data Availability Statement}: 
All data generated or analyzed during this study are available within the paper and its Supplementary Information. 
Further source data will be made available online once published.

\end{footnotesize}

% \vspace{-0.3cm}
\bibliographystyle{apsrev4-2}
\bibliography{bibliography}

\begin{thebibliography}{61}%
\makeatletter
\providecommand \@ifxundefined [1]{%
 \@ifx{#1\undefined}
}%
\providecommand \@ifnum [1]{%
 \ifnum #1\expandafter \@firstoftwo
 \else \expandafter \@secondoftwo
 \fi
}%
\providecommand \@ifx [1]{%
 \ifx #1\expandafter \@firstoftwo
 \else \expandafter \@secondoftwo
 \fi
}%
\providecommand \natexlab [1]{#1}%
\providecommand \emph  [1]{``#1''}%
\providecommand \bibnamefont  [1]{#1}%
\providecommand \bibfnamefont [1]{#1}%
\providecommand \citenamefont [1]{#1}%
\providecommand \href@noop [0]{\@secondoftwo}%
\providecommand \href [0]{\begingroup \@sanitize@url \@href}%
\providecommand \@href[1]{\@@startlink{#1}\@@href}%
\providecommand \@@href[1]{\endgroup#1\@@endlink}%
\providecommand \@sanitize@url [0]{\catcode `\\12\catcode `\$12\catcode
  `\&12\catcode `\#12\catcode `\^12\catcode `\_12\catcode `\%12\relax}%
\providecommand \@@startlink[1]{}%
\providecommand \@@endlink[0]{}%
\providecommand \url  [0]{\begingroup\@sanitize@url \@url }%
\providecommand \@url [1]{\endgroup\@href {#1}{\urlprefix }}%
\providecommand \urlprefix  [0]{URL }%
\providecommand \Eprint [0]{\href }%
\providecommand \doibase [0]{https://doi.org/}%
\providecommand \selectlanguage [0]{\@gobble}%
\providecommand \bibinfo  [0]{\@secondoftwo}%
\providecommand \bibfield  [0]{\@secondoftwo}%
\providecommand \translation [1]{[#1]}%
\providecommand \BibitemOpen [0]{}%
\providecommand \bibitemStop [0]{}%
\providecommand \bibitemNoStop [0]{.\EOS\space}%
\providecommand \EOS [0]{\spacefactor3000\relax}%
\providecommand \BibitemShut  [1]{\csname bibitem#1\endcsname}%
\let\auto@bib@innerbib\@empty
%</preamble>
\bibitem [{\citenamefont {Margalit}\ \emph {et~al.}(2021)\citenamefont
  {Margalit}, \citenamefont {Xiang}, \citenamefont {Bowers}, \citenamefont
  {Bjorlin}, \citenamefont {Blum},\ and\ \citenamefont
  {Bowers}}]{margalit2021perspective}%
  \BibitemOpen
  \bibfield  {author} {\bibinfo {author} {\bibfnamefont {N.}~\bibnamefont
  {Margalit}}, \bibinfo {author} {\bibfnamefont {C.}~\bibnamefont {Xiang}},
  \bibinfo {author} {\bibfnamefont {S.~M.}\ \bibnamefont {Bowers}}, \bibinfo
  {author} {\bibfnamefont {A.}~\bibnamefont {Bjorlin}}, \bibinfo {author}
  {\bibfnamefont {R.}~\bibnamefont {Blum}},\ and\ \bibinfo {author}
  {\bibfnamefont {J.~E.}\ \bibnamefont {Bowers}},\ }\bibfield  {title}
  {\bibinfo {title} {Perspective on the future of silicon photonics and
  electronics},\ }\href {https://aip.scitation.org/doi/full/10.1063/5.0050117}
  {\bibfield  {journal} {\bibinfo  {journal} {Applied Physics Letters}\
  }\textbf {\bibinfo {volume} {118}},\ \bibinfo {pages} {220501} (\bibinfo
  {year} {2021})}\BibitemShut {NoStop}%
\bibitem [{\citenamefont {Doerr}(2015)}]{doerr2015silicon}%
  \BibitemOpen
  \bibfield  {author} {\bibinfo {author} {\bibfnamefont {C.}~\bibnamefont
  {Doerr}},\ }\bibfield  {title} {\bibinfo {title} {Silicon photonic
  integration in telecommunications},\ }\href
  {https://www.frontiersin.org/articles/10.3389/fphy.2015.00037} {\bibfield
  {journal} {\bibinfo  {journal} {Frontiers in Physics}\ }\textbf {\bibinfo
  {volume} {3}} (\bibinfo {year} {2015})}\BibitemShut {NoStop}%
\bibitem [{\citenamefont {Thomson}\ \emph {et~al.}(2016)\citenamefont
  {Thomson}, \citenamefont {Zilkie}, \citenamefont {Bowers}, \citenamefont
  {Komljenovic}, \citenamefont {Reed}, \citenamefont {Vivien}, \citenamefont
  {Marris-Morini}, \citenamefont {Cassan}, \citenamefont {Virot}, \citenamefont
  {F{\'{e}}d{\'{e}}li}, \citenamefont {Hartmann}, \citenamefont {Schmid},
  \citenamefont {Xu}, \citenamefont {Boeuf}, \citenamefont {O'Brien},
  \citenamefont {Mashanovich},\ and\ \citenamefont {Nedeljkovic}}]{Thomson:16}%
  \BibitemOpen
  \bibfield  {author} {\bibinfo {author} {\bibfnamefont {D.}~\bibnamefont
  {Thomson}}, \bibinfo {author} {\bibfnamefont {A.}~\bibnamefont {Zilkie}},
  \bibinfo {author} {\bibfnamefont {J.~E.}\ \bibnamefont {Bowers}}, \bibinfo
  {author} {\bibfnamefont {T.}~\bibnamefont {Komljenovic}}, \bibinfo {author}
  {\bibfnamefont {G.~T.}\ \bibnamefont {Reed}}, \bibinfo {author}
  {\bibfnamefont {L.}~\bibnamefont {Vivien}}, \bibinfo {author} {\bibfnamefont
  {D.}~\bibnamefont {Marris-Morini}}, \bibinfo {author} {\bibfnamefont
  {E.}~\bibnamefont {Cassan}}, \bibinfo {author} {\bibfnamefont
  {L.}~\bibnamefont {Virot}}, \bibinfo {author} {\bibfnamefont {J.-M.}\
  \bibnamefont {F{\'{e}}d{\'{e}}li}}, \bibinfo {author} {\bibfnamefont {J.-M.}\
  \bibnamefont {Hartmann}}, \bibinfo {author} {\bibfnamefont {J.~H.}\
  \bibnamefont {Schmid}}, \bibinfo {author} {\bibfnamefont {D.-X.}\
  \bibnamefont {Xu}}, \bibinfo {author} {\bibfnamefont {F.}~\bibnamefont
  {Boeuf}}, \bibinfo {author} {\bibfnamefont {P.}~\bibnamefont {O'Brien}},
  \bibinfo {author} {\bibfnamefont {G.~Z.}\ \bibnamefont {Mashanovich}},\ and\
  \bibinfo {author} {\bibfnamefont {M.}~\bibnamefont {Nedeljkovic}},\
  }\bibfield  {title} {\bibinfo {title} {Roadmap on silicon photonics},\ }\href
  {https://doi.org/10.1088/2040-8978/18/7/073003} {\bibfield  {journal}
  {\bibinfo  {journal} {Journal of Optics}\ }\textbf {\bibinfo {volume} {18}},\
  \bibinfo {pages} {073003} (\bibinfo {year} {2016})}\BibitemShut {NoStop}%
\bibitem [{\citenamefont {Cheng}\ \emph {et~al.}(2018)\citenamefont {Cheng},
  \citenamefont {Bahadori}, \citenamefont {Glick}, \citenamefont {Rumley},\
  and\ \citenamefont {Bergman}}]{Cheng:18}%
  \BibitemOpen
  \bibfield  {author} {\bibinfo {author} {\bibfnamefont {Q.}~\bibnamefont
  {Cheng}}, \bibinfo {author} {\bibfnamefont {M.}~\bibnamefont {Bahadori}},
  \bibinfo {author} {\bibfnamefont {M.}~\bibnamefont {Glick}}, \bibinfo
  {author} {\bibfnamefont {S.}~\bibnamefont {Rumley}},\ and\ \bibinfo {author}
  {\bibfnamefont {K.}~\bibnamefont {Bergman}},\ }\bibfield  {title} {\bibinfo
  {title} {Recent advances in optical technologies for data centers: a
  review},\ }\href {https://doi.org/10.1364/OPTICA.5.001354} {\bibfield
  {journal} {\bibinfo  {journal} {Optica}\ }\textbf {\bibinfo {volume} {5}},\
  \bibinfo {pages} {1354} (\bibinfo {year} {2018})}\BibitemShut {NoStop}%
\bibitem [{\citenamefont {Soref}(2006)}]{Soref:06}%
  \BibitemOpen
  \bibfield  {author} {\bibinfo {author} {\bibfnamefont {R.}~\bibnamefont
  {Soref}},\ }\bibfield  {title} {\bibinfo {title} {The Past, Present, and
  Future of Silicon Photonics},\ }\href
  {https://doi.org/10.1109/JSTQE.2006.883151} {\bibfield  {journal} {\bibinfo
  {journal} {IEEE Journal of Selected Topics in Quantum Electronics}\ }\textbf
  {\bibinfo {volume} {12}},\ \bibinfo {pages} {1678} (\bibinfo {year}
  {2006})}\BibitemShut {NoStop}%
\bibitem [{\citenamefont {Marpaung}\ \emph {et~al.}(2019)\citenamefont
  {Marpaung}, \citenamefont {Yao},\ and\ \citenamefont
  {Capmany}}]{Marpaung:19}%
  \BibitemOpen
  \bibfield  {author} {\bibinfo {author} {\bibfnamefont {D.}~\bibnamefont
  {Marpaung}}, \bibinfo {author} {\bibfnamefont {J.}~\bibnamefont {Yao}},\ and\
  \bibinfo {author} {\bibfnamefont {J.}~\bibnamefont {Capmany}},\ }\bibfield
  {title} {\bibinfo {title} {Integrated microwave photonics},\ }\href
  {https://doi.org/10.1038/s41566-018-0310-5} {\bibfield  {journal} {\bibinfo
  {journal} {Nature Photonics}\ }\textbf {\bibinfo {volume} {13}},\ \bibinfo
  {pages} {80} (\bibinfo {year} {2019})}\BibitemShut {NoStop}%
\bibitem [{\citenamefont {Lefevre}(2014)}]{lefevre2014fiber}%
  \BibitemOpen
  \bibfield  {author} {\bibinfo {author} {\bibfnamefont {H.~C.}\ \bibnamefont
  {Lefevre}},\ }\href@noop {} {\emph {\bibinfo {title} {The fiber-optic
  gyroscope}}}\ (\bibinfo  {publisher} {Artech house},\ \bibinfo {year}
  {2014})\BibitemShut {NoStop}%
\bibitem [{\citenamefont {Ludlow}\ \emph {et~al.}(2015)\citenamefont {Ludlow},
  \citenamefont {Boyd}, \citenamefont {Ye}, \citenamefont {Peik},\ and\
  \citenamefont {Schmidt}}]{ludlow2015optical}%
  \BibitemOpen
  \bibfield  {author} {\bibinfo {author} {\bibfnamefont {A.~D.}\ \bibnamefont
  {Ludlow}}, \bibinfo {author} {\bibfnamefont {M.~M.}\ \bibnamefont {Boyd}},
  \bibinfo {author} {\bibfnamefont {J.}~\bibnamefont {Ye}}, \bibinfo {author}
  {\bibfnamefont {E.}~\bibnamefont {Peik}},\ and\ \bibinfo {author}
  {\bibfnamefont {P.~O.}\ \bibnamefont {Schmidt}},\ }\bibfield  {title}
  {\bibinfo {title} {Optical atomic clocks},\ }\href
  {https://journals.aps.org/rmp/abstract/10.1103/RevModPhys.87.637} {\bibfield
  {journal} {\bibinfo  {journal} {Reviews of Modern Physics}\ }\textbf
  {\bibinfo {volume} {87}},\ \bibinfo {pages} {637} (\bibinfo {year}
  {2015})}\BibitemShut {NoStop}%
\bibitem [{\citenamefont {Siew}\ \emph {et~al.}(2021)\citenamefont {Siew},
  \citenamefont {Li}, \citenamefont {Gao}, \citenamefont {Zheng}, \citenamefont
  {Zhang}, \citenamefont {Guo}, \citenamefont {Xie}, \citenamefont {Song},
  \citenamefont {Dong}, \citenamefont {Luo}, \citenamefont {Li}, \citenamefont
  {Luo},\ and\ \citenamefont {Lo}}]{siew2021review}%
  \BibitemOpen
  \bibfield  {author} {\bibinfo {author} {\bibfnamefont {S.~Y.}\ \bibnamefont
  {Siew}}, \bibinfo {author} {\bibfnamefont {B.}~\bibnamefont {Li}}, \bibinfo
  {author} {\bibfnamefont {F.}~\bibnamefont {Gao}}, \bibinfo {author}
  {\bibfnamefont {H.~Y.}\ \bibnamefont {Zheng}}, \bibinfo {author}
  {\bibfnamefont {W.}~\bibnamefont {Zhang}}, \bibinfo {author} {\bibfnamefont
  {P.}~\bibnamefont {Guo}}, \bibinfo {author} {\bibfnamefont {S.~W.}\
  \bibnamefont {Xie}}, \bibinfo {author} {\bibfnamefont {A.}~\bibnamefont
  {Song}}, \bibinfo {author} {\bibfnamefont {B.}~\bibnamefont {Dong}}, \bibinfo
  {author} {\bibfnamefont {L.~W.}\ \bibnamefont {Luo}}, \bibinfo {author}
  {\bibfnamefont {C.}~\bibnamefont {Li}}, \bibinfo {author} {\bibfnamefont
  {X.}~\bibnamefont {Luo}},\ and\ \bibinfo {author} {\bibfnamefont {G.-Q.}\
  \bibnamefont {Lo}},\ }\bibfield  {title} {\bibinfo {title} {Review of Silicon
  Photonics Technology and Platform Development},\ }\href
  {https://doi.org/10.1109/JLT.2021.3066203} {\bibfield  {journal} {\bibinfo
  {journal} {Journal of Lightwave Technology}\ }\textbf {\bibinfo {volume}
  {39}},\ \bibinfo {pages} {4374} (\bibinfo {year} {2021})}\BibitemShut
  {NoStop}%
\bibitem [{\citenamefont {Giewont}\ \emph {et~al.}(2019)\citenamefont
  {Giewont}, \citenamefont {Nummy}, \citenamefont {Anderson}, \citenamefont
  {Ayala}, \citenamefont {Barwicz}, \citenamefont {Bian}, \citenamefont
  {Dezfulian}, \citenamefont {Gill}, \citenamefont {Houghton}, \citenamefont
  {Hu}, \citenamefont {Peng}, \citenamefont {Rakowski}, \citenamefont {Rauch},
  \citenamefont {Rosenberg}, \citenamefont {Sahin}, \citenamefont {Stobert},\
  and\ \citenamefont {Stricker}}]{Giewont2019300mm}%
  \BibitemOpen
  \bibfield  {author} {\bibinfo {author} {\bibfnamefont {K.}~\bibnamefont
  {Giewont}}, \bibinfo {author} {\bibfnamefont {K.}~\bibnamefont {Nummy}},
  \bibinfo {author} {\bibfnamefont {F.~A.}\ \bibnamefont {Anderson}}, \bibinfo
  {author} {\bibfnamefont {J.}~\bibnamefont {Ayala}}, \bibinfo {author}
  {\bibfnamefont {T.}~\bibnamefont {Barwicz}}, \bibinfo {author} {\bibfnamefont
  {Y.}~\bibnamefont {Bian}}, \bibinfo {author} {\bibfnamefont {K.~K.}\
  \bibnamefont {Dezfulian}}, \bibinfo {author} {\bibfnamefont {D.~M.}\
  \bibnamefont {Gill}}, \bibinfo {author} {\bibfnamefont {T.}~\bibnamefont
  {Houghton}}, \bibinfo {author} {\bibfnamefont {S.}~\bibnamefont {Hu}},
  \bibinfo {author} {\bibfnamefont {B.}~\bibnamefont {Peng}}, \bibinfo {author}
  {\bibfnamefont {M.}~\bibnamefont {Rakowski}}, \bibinfo {author}
  {\bibfnamefont {S.}~\bibnamefont {Rauch}}, \bibinfo {author} {\bibfnamefont
  {J.~C.}\ \bibnamefont {Rosenberg}}, \bibinfo {author} {\bibfnamefont
  {A.}~\bibnamefont {Sahin}}, \bibinfo {author} {\bibfnamefont
  {I.}~\bibnamefont {Stobert}},\ and\ \bibinfo {author} {\bibfnamefont
  {A.}~\bibnamefont {Stricker}},\ }\bibfield  {title} {\bibinfo {title} {300-mm
  Monolithic Silicon Photonics Foundry Technology},\ }\href
  {https://doi.org/10.1109/JSTQE.2019.2908790} {\bibfield  {journal} {\bibinfo
  {journal} {IEEE Journal of Selected Topics in Quantum Electronics}\ }\textbf
  {\bibinfo {volume} {25}},\ \bibinfo {pages} {1} (\bibinfo {year}
  {2019})}\BibitemShut {NoStop}%
\bibitem [{\citenamefont {Rahim}\ \emph {et~al.}(2019)\citenamefont {Rahim},
  \citenamefont {Goyvaerts}, \citenamefont {Szelag}, \citenamefont {Fedeli},
  \citenamefont {Absil}, \citenamefont {Aalto}, \citenamefont {Harjanne},
  \citenamefont {Littlejohns}, \citenamefont {Reed}, \citenamefont {Winzer},
  \citenamefont {Lischke}, \citenamefont {Zimmermann}, \citenamefont {Knoll},
  \citenamefont {Geuzebroek}, \citenamefont {Leinse}, \citenamefont
  {Geiselmann}, \citenamefont {Zervas}, \citenamefont {Jans}, \citenamefont
  {Stassen}, \citenamefont {Domínguez}, \citenamefont {Muñoz}, \citenamefont
  {Domenech}, \citenamefont {Giesecke}, \citenamefont {Lemme},\ and\
  \citenamefont {Baets}}]{rahim2019open}%
  \BibitemOpen
  \bibfield  {author} {\bibinfo {author} {\bibfnamefont {A.}~\bibnamefont
  {Rahim}}, \bibinfo {author} {\bibfnamefont {J.}~\bibnamefont {Goyvaerts}},
  \bibinfo {author} {\bibfnamefont {B.}~\bibnamefont {Szelag}}, \bibinfo
  {author} {\bibfnamefont {J.-M.}\ \bibnamefont {Fedeli}}, \bibinfo {author}
  {\bibfnamefont {P.}~\bibnamefont {Absil}}, \bibinfo {author} {\bibfnamefont
  {T.}~\bibnamefont {Aalto}}, \bibinfo {author} {\bibfnamefont
  {M.}~\bibnamefont {Harjanne}}, \bibinfo {author} {\bibfnamefont
  {C.}~\bibnamefont {Littlejohns}}, \bibinfo {author} {\bibfnamefont
  {G.}~\bibnamefont {Reed}}, \bibinfo {author} {\bibfnamefont {G.}~\bibnamefont
  {Winzer}}, \bibinfo {author} {\bibfnamefont {S.}~\bibnamefont {Lischke}},
  \bibinfo {author} {\bibfnamefont {L.}~\bibnamefont {Zimmermann}}, \bibinfo
  {author} {\bibfnamefont {D.}~\bibnamefont {Knoll}}, \bibinfo {author}
  {\bibfnamefont {D.}~\bibnamefont {Geuzebroek}}, \bibinfo {author}
  {\bibfnamefont {A.}~\bibnamefont {Leinse}}, \bibinfo {author} {\bibfnamefont
  {M.}~\bibnamefont {Geiselmann}}, \bibinfo {author} {\bibfnamefont
  {M.}~\bibnamefont {Zervas}}, \bibinfo {author} {\bibfnamefont
  {H.}~\bibnamefont {Jans}}, \bibinfo {author} {\bibfnamefont {A.}~\bibnamefont
  {Stassen}}, \bibinfo {author} {\bibfnamefont {C.}~\bibnamefont {Domínguez}},
  \bibinfo {author} {\bibfnamefont {P.}~\bibnamefont {Muñoz}}, \bibinfo
  {author} {\bibfnamefont {D.}~\bibnamefont {Domenech}}, \bibinfo {author}
  {\bibfnamefont {A.~L.}\ \bibnamefont {Giesecke}}, \bibinfo {author}
  {\bibfnamefont {M.~C.}\ \bibnamefont {Lemme}},\ and\ \bibinfo {author}
  {\bibfnamefont {R.}~\bibnamefont {Baets}},\ }\bibfield  {title} {\bibinfo
  {title} {Open-Access Silicon Photonics Platforms in Europe},\ }\href
  {https://doi.org/10.1109/JSTQE.2019.2915949} {\bibfield  {journal} {\bibinfo
  {journal} {IEEE Journal of Selected Topics in Quantum Electronics}\ }\textbf
  {\bibinfo {volume} {25}},\ \bibinfo {pages} {1} (\bibinfo {year}
  {2019})}\BibitemShut {NoStop}%
\bibitem [{\citenamefont {Coldren}\ \emph {et~al.}(2012)\citenamefont
  {Coldren}, \citenamefont {Corzine},\ and\ \citenamefont
  {Mashanovitch}}]{coldren_diode_2012}%
  \BibitemOpen
  \bibfield  {author} {\bibinfo {author} {\bibfnamefont {L.~A.}\ \bibnamefont
  {Coldren}}, \bibinfo {author} {\bibfnamefont {S.~W.}\ \bibnamefont
  {Corzine}},\ and\ \bibinfo {author} {\bibfnamefont {M.}~\bibnamefont
  {Mashanovitch}},\ }\href
  {https://onlinelibrary.wiley.com/doi/book/10.1002/9781118148167} {\emph
  {\bibinfo {title} {Diode lasers and photonic integrated circuits}}}\
  (\bibinfo  {publisher} {Wiley},\ \bibinfo {year} {2012})\BibitemShut
  {NoStop}%
\bibitem [{\citenamefont {Henry}(1982)}]{henry1982theory}%
  \BibitemOpen
  \bibfield  {author} {\bibinfo {author} {\bibfnamefont {C.}~\bibnamefont
  {Henry}},\ }\bibfield  {title} {\bibinfo {title} {Theory of the linewidth of
  semiconductor lasers},\ }\href {https://doi.org/10.1109/JQE.1982.1071522}
  {\bibfield  {journal} {\bibinfo  {journal} {IEEE Journal of Quantum
  Electronics}\ }\textbf {\bibinfo {volume} {18}},\ \bibinfo {pages} {259}
  (\bibinfo {year} {1982})}\BibitemShut {NoStop}%
\bibitem [{\citenamefont {Tkach}\ and\ \citenamefont
  {Chraplyvy}(1986)}]{Tkach1986regimes}%
  \BibitemOpen
  \bibfield  {author} {\bibinfo {author} {\bibfnamefont {R.}~\bibnamefont
  {Tkach}}\ and\ \bibinfo {author} {\bibfnamefont {A.}~\bibnamefont
  {Chraplyvy}},\ }\bibfield  {title} {\bibinfo {title} {Regimes of feedback
  effects in 1.5-µm distributed feedback lasers},\ }\href
  {https://doi.org/10.1109/JLT.1986.1074666} {\bibfield  {journal} {\bibinfo
  {journal} {Journal of Lightwave Technology}\ }\textbf {\bibinfo {volume}
  {4}},\ \bibinfo {pages} {1655} (\bibinfo {year} {1986})}\BibitemShut
  {NoStop}%
\bibitem [{\citenamefont {Carroll}\ \emph {et~al.}(2016)\citenamefont
  {Carroll}, \citenamefont {Lee}, \citenamefont {Scarcella}, \citenamefont
  {Gradkowski}, \citenamefont {Duperron}, \citenamefont {Lu}, \citenamefont
  {Zhao}, \citenamefont {Eason}, \citenamefont {Morrissey}, \citenamefont
  {Rensing}, \citenamefont {Collins}, \citenamefont {Hwang},\ and\
  \citenamefont {O’Brien}}]{Lee2016photonic}%
  \BibitemOpen
  \bibfield  {author} {\bibinfo {author} {\bibfnamefont {L.}~\bibnamefont
  {Carroll}}, \bibinfo {author} {\bibfnamefont {J.-S.}\ \bibnamefont {Lee}},
  \bibinfo {author} {\bibfnamefont {C.}~\bibnamefont {Scarcella}}, \bibinfo
  {author} {\bibfnamefont {K.}~\bibnamefont {Gradkowski}}, \bibinfo {author}
  {\bibfnamefont {M.}~\bibnamefont {Duperron}}, \bibinfo {author}
  {\bibfnamefont {H.}~\bibnamefont {Lu}}, \bibinfo {author} {\bibfnamefont
  {Y.}~\bibnamefont {Zhao}}, \bibinfo {author} {\bibfnamefont {C.}~\bibnamefont
  {Eason}}, \bibinfo {author} {\bibfnamefont {P.}~\bibnamefont {Morrissey}},
  \bibinfo {author} {\bibfnamefont {M.}~\bibnamefont {Rensing}}, \bibinfo
  {author} {\bibfnamefont {S.}~\bibnamefont {Collins}}, \bibinfo {author}
  {\bibfnamefont {H.~Y.}\ \bibnamefont {Hwang}},\ and\ \bibinfo {author}
  {\bibfnamefont {P.}~\bibnamefont {O’Brien}},\ }\bibfield  {title} {\bibinfo
  {title} {Photonic Packaging: Transforming Silicon Photonic Integrated
  Circuits into Photonic Devices},\ }\href
  {https://www.mdpi.com/2076-3417/6/12/426} {\bibfield  {journal} {\bibinfo
  {journal} {Applied Sciences}\ }\textbf {\bibinfo {volume} {6}} (\bibinfo
  {year} {2016})}\BibitemShut {NoStop}%
\bibitem [{\citenamefont {Xiang}\ \emph
  {et~al.}(2022{\natexlab{a}})\citenamefont {Xiang}, \citenamefont {Jin},
  \citenamefont {Huang}, \citenamefont {Tran}, \citenamefont {Guo},
  \citenamefont {Wan}, \citenamefont {Xie}, \citenamefont {Kurczveil},
  \citenamefont {Netherton}, \citenamefont {Liang}, \citenamefont {Rong},\ and\
  \citenamefont {Bowers}}]{xiang2021JSTQE}%
  \BibitemOpen
  \bibfield  {author} {\bibinfo {author} {\bibfnamefont {C.}~\bibnamefont
  {Xiang}}, \bibinfo {author} {\bibfnamefont {W.}~\bibnamefont {Jin}}, \bibinfo
  {author} {\bibfnamefont {D.}~\bibnamefont {Huang}}, \bibinfo {author}
  {\bibfnamefont {M.~A.}\ \bibnamefont {Tran}}, \bibinfo {author}
  {\bibfnamefont {J.}~\bibnamefont {Guo}}, \bibinfo {author} {\bibfnamefont
  {Y.}~\bibnamefont {Wan}}, \bibinfo {author} {\bibfnamefont {W.}~\bibnamefont
  {Xie}}, \bibinfo {author} {\bibfnamefont {G.}~\bibnamefont {Kurczveil}},
  \bibinfo {author} {\bibfnamefont {A.~M.}\ \bibnamefont {Netherton}}, \bibinfo
  {author} {\bibfnamefont {D.}~\bibnamefont {Liang}}, \bibinfo {author}
  {\bibfnamefont {H.}~\bibnamefont {Rong}},\ and\ \bibinfo {author}
  {\bibfnamefont {J.~E.}\ \bibnamefont {Bowers}},\ }\bibfield  {title}
  {\bibinfo {title} {High-Performance Silicon Photonics Using Heterogeneous
  Integration},\ }\href {https://doi.org/10.1109/JSTQE.2021.3126124} {\bibfield
   {journal} {\bibinfo  {journal} {IEEE Journal of Selected Topics in Quantum
  Electronics}\ }\textbf {\bibinfo {volume} {28}},\ \bibinfo {pages} {1}
  (\bibinfo {year} {2022}{\natexlab{a}})}\BibitemShut {NoStop}%
\bibitem [{\citenamefont {Liang}\ and\ \citenamefont
  {Bowers}(2021)}]{liang2021recent}%
  \BibitemOpen
  \bibfield  {author} {\bibinfo {author} {\bibfnamefont {D.}~\bibnamefont
  {Liang}}\ and\ \bibinfo {author} {\bibfnamefont {J.~E.}\ \bibnamefont
  {Bowers}},\ }\bibfield  {title} {\bibinfo {title} {Recent progress in
  heterogeneous {III-V}-on-silicon photonic integration},\ }\href
  {https://light-am.com/en/article/doi/10.37188/lam.2021.005} {\bibfield
  {journal} {\bibinfo  {journal} {Light: Advanced Manufacturing}\ }\textbf
  {\bibinfo {volume} {2}},\ \bibinfo {pages} {59} (\bibinfo {year}
  {2021})}\BibitemShut {NoStop}%
\bibitem [{\citenamefont {Marshall}\ \emph {et~al.}(2018)\citenamefont
  {Marshall}, \citenamefont {Hsu}, \citenamefont {Wang}, \citenamefont
  {Kunert}, \citenamefont {Koos},\ and\ \citenamefont
  {Van~Thourhout}}]{marshall2018heterogeneous}%
  \BibitemOpen
  \bibfield  {author} {\bibinfo {author} {\bibfnamefont {O.}~\bibnamefont
  {Marshall}}, \bibinfo {author} {\bibfnamefont {M.}~\bibnamefont {Hsu}},
  \bibinfo {author} {\bibfnamefont {Z.}~\bibnamefont {Wang}}, \bibinfo {author}
  {\bibfnamefont {B.}~\bibnamefont {Kunert}}, \bibinfo {author} {\bibfnamefont
  {C.}~\bibnamefont {Koos}},\ and\ \bibinfo {author} {\bibfnamefont
  {D.}~\bibnamefont {Van~Thourhout}},\ }\bibfield  {title} {\bibinfo {title}
  {Heterogeneous integration on silicon photonics},\ }\href
  {https://ieeexplore.ieee.org/document/6599998} {\bibfield  {journal}
  {\bibinfo  {journal} {Proceedings of the IEEE}\ }\textbf {\bibinfo {volume}
  {106}},\ \bibinfo {pages} {2258} (\bibinfo {year} {2018})}\BibitemShut
  {NoStop}%
\bibitem [{\citenamefont {Pintus}\ \emph {et~al.}(2017)\citenamefont {Pintus},
  \citenamefont {Huang}, \citenamefont {Zhang}, \citenamefont {Shoji},
  \citenamefont {Mizumoto},\ and\ \citenamefont
  {Bowers}}]{pintus2017microring}%
  \BibitemOpen
  \bibfield  {author} {\bibinfo {author} {\bibfnamefont {P.}~\bibnamefont
  {Pintus}}, \bibinfo {author} {\bibfnamefont {D.}~\bibnamefont {Huang}},
  \bibinfo {author} {\bibfnamefont {C.}~\bibnamefont {Zhang}}, \bibinfo
  {author} {\bibfnamefont {Y.}~\bibnamefont {Shoji}}, \bibinfo {author}
  {\bibfnamefont {T.}~\bibnamefont {Mizumoto}},\ and\ \bibinfo {author}
  {\bibfnamefont {J.~E.}\ \bibnamefont {Bowers}},\ }\bibfield  {title}
  {\bibinfo {title} {Microring-Based Optical Isolator and Circulator with
  Integrated Electromagnet for Silicon Photonics},\ }\href
  {https://doi.org/10.1109/JLT.2016.2644626} {\bibfield  {journal} {\bibinfo
  {journal} {Journal of Lightwave Technology}\ }\textbf {\bibinfo {volume}
  {35}},\ \bibinfo {pages} {1429} (\bibinfo {year} {2017})}\BibitemShut
  {NoStop}%
\bibitem [{\citenamefont {Santis}\ \emph {et~al.}(2014)\citenamefont {Santis},
  \citenamefont {Steger}, \citenamefont {Vilenchik}, \citenamefont {Vasilyev},\
  and\ \citenamefont {Yariv}}]{santis2014high}%
  \BibitemOpen
  \bibfield  {author} {\bibinfo {author} {\bibfnamefont {C.~T.}\ \bibnamefont
  {Santis}}, \bibinfo {author} {\bibfnamefont {S.~T.}\ \bibnamefont {Steger}},
  \bibinfo {author} {\bibfnamefont {Y.}~\bibnamefont {Vilenchik}}, \bibinfo
  {author} {\bibfnamefont {A.}~\bibnamefont {Vasilyev}},\ and\ \bibinfo
  {author} {\bibfnamefont {A.}~\bibnamefont {Yariv}},\ }\bibfield  {title}
  {\bibinfo {title} {High-coherence semiconductor lasers based on integral
  high-{Q} resonators in hybrid {Si/III-V} platforms},\ }\href
  {https://www.pnas.org/content/111/8/2879} {\bibfield  {journal} {\bibinfo
  {journal} {Proceedings of the National Academy of Sciences}\ }\textbf
  {\bibinfo {volume} {111}},\ \bibinfo {pages} {2879} (\bibinfo {year}
  {2014})}\BibitemShut {NoStop}%
\bibitem [{\citenamefont {Harfouche}\ \emph {et~al.}(2020)\citenamefont
  {Harfouche}, \citenamefont {Kim}, \citenamefont {Wang}, \citenamefont
  {Santis}, \citenamefont {Zhang}, \citenamefont {Chen}, \citenamefont
  {Satyan}, \citenamefont {Rakuljic},\ and\ \citenamefont
  {Yariv}}]{Harfouche:20}%
  \BibitemOpen
  \bibfield  {author} {\bibinfo {author} {\bibfnamefont {M.}~\bibnamefont
  {Harfouche}}, \bibinfo {author} {\bibfnamefont {D.}~\bibnamefont {Kim}},
  \bibinfo {author} {\bibfnamefont {H.}~\bibnamefont {Wang}}, \bibinfo {author}
  {\bibfnamefont {C.~T.}\ \bibnamefont {Santis}}, \bibinfo {author}
  {\bibfnamefont {Z.}~\bibnamefont {Zhang}}, \bibinfo {author} {\bibfnamefont
  {H.}~\bibnamefont {Chen}}, \bibinfo {author} {\bibfnamefont {N.}~\bibnamefont
  {Satyan}}, \bibinfo {author} {\bibfnamefont {G.}~\bibnamefont {Rakuljic}},\
  and\ \bibinfo {author} {\bibfnamefont {A.}~\bibnamefont {Yariv}},\ }\bibfield
   {title} {\bibinfo {title} {Kicking the habit/semiconductor lasers without
  isolators},\ }\href {https://doi.org/10.1364/OE.411816} {\bibfield  {journal}
  {\bibinfo  {journal} {Opt. Express}\ }\textbf {\bibinfo {volume} {28}},\
  \bibinfo {pages} {36466} (\bibinfo {year} {2020})}\BibitemShut {NoStop}%
\bibitem [{\citenamefont {Kippenberg}\ \emph {et~al.}(2018)\citenamefont
  {Kippenberg}, \citenamefont {Gaeta}, \citenamefont {Lipson},\ and\
  \citenamefont {Gorodetsky}}]{Kippenberg:18}%
  \BibitemOpen
  \bibfield  {author} {\bibinfo {author} {\bibfnamefont {T.~J.}\ \bibnamefont
  {Kippenberg}}, \bibinfo {author} {\bibfnamefont {A.~L.}\ \bibnamefont
  {Gaeta}}, \bibinfo {author} {\bibfnamefont {M.}~\bibnamefont {Lipson}},\ and\
  \bibinfo {author} {\bibfnamefont {M.~L.}\ \bibnamefont {Gorodetsky}},\
  }\bibfield  {title} {\bibinfo {title} {Dissipative Kerr solitons in optical
  microresonators},\ }\href
  {http://science.sciencemag.org/content/361/6402/eaan8083} {\bibfield
  {journal} {\bibinfo  {journal} {Science}\ }\textbf {\bibinfo {volume} {361}}
  (\bibinfo {year} {2018})}\BibitemShut {NoStop}%
\bibitem [{\citenamefont {Gaeta}\ \emph {et~al.}(2019)\citenamefont {Gaeta},
  \citenamefont {Lipson},\ and\ \citenamefont {Kippenberg}}]{Gaeta:19}%
  \BibitemOpen
  \bibfield  {author} {\bibinfo {author} {\bibfnamefont {A.~L.}\ \bibnamefont
  {Gaeta}}, \bibinfo {author} {\bibfnamefont {M.}~\bibnamefont {Lipson}},\ and\
  \bibinfo {author} {\bibfnamefont {T.~J.}\ \bibnamefont {Kippenberg}},\
  }\bibfield  {title} {\bibinfo {title} {Photonic-chip-based frequency combs},\
  }\href {https://doi.org/10.1038/s41566-019-0358-x} {\bibfield  {journal}
  {\bibinfo  {journal} {Nature Photonics}\ }\textbf {\bibinfo {volume} {13}},\
  \bibinfo {pages} {158} (\bibinfo {year} {2019})}\BibitemShut {NoStop}%
\bibitem [{\citenamefont {Xiang}\ \emph
  {et~al.}(2022{\natexlab{b}})\citenamefont {Xiang}, \citenamefont {Jin},\ and\
  \citenamefont {Bowers}}]{Xiang2022silicon}%
  \BibitemOpen
  \bibfield  {author} {\bibinfo {author} {\bibfnamefont {C.}~\bibnamefont
  {Xiang}}, \bibinfo {author} {\bibfnamefont {W.}~\bibnamefont {Jin}},\ and\
  \bibinfo {author} {\bibfnamefont {J.~E.}\ \bibnamefont {Bowers}},\ }\bibfield
   {title} {\bibinfo {title} {Silicon nitride passive and active photonic
  integrated circuits: trends and prospects},\ }\href
  {https://doi.org/10.1364/PRJ.452936} {\bibfield  {journal} {\bibinfo
  {journal} {Photon. Res.}\ }\textbf {\bibinfo {volume} {10}},\ \bibinfo
  {pages} {A82} (\bibinfo {year} {2022}{\natexlab{b}})}\BibitemShut {NoStop}%
\bibitem [{\citenamefont {Dai}\ \emph {et~al.}(2012)\citenamefont {Dai},
  \citenamefont {Bauters},\ and\ \citenamefont {Bowers}}]{Dai2012passive}%
  \BibitemOpen
  \bibfield  {author} {\bibinfo {author} {\bibfnamefont {D.}~\bibnamefont
  {Dai}}, \bibinfo {author} {\bibfnamefont {J.}~\bibnamefont {Bauters}},\ and\
  \bibinfo {author} {\bibfnamefont {J.~E.}\ \bibnamefont {Bowers}},\ }\bibfield
   {title} {\bibinfo {title} {Passive technologies for future large-scale
  photonic integrated circuits on silicon: polarization handling, light
  non-reciprocity and loss reduction},\ }\href
  {https://doi.org/10.1038/lsa.2012.1} {\bibfield  {journal} {\bibinfo
  {journal} {Light: Science {\&} Applications}\ }\textbf {\bibinfo {volume}
  {1}},\ \bibinfo {pages} {e1} (\bibinfo {year} {2012})}\BibitemShut {NoStop}%
\bibitem [{\citenamefont {Bauters}\ \emph {et~al.}(2011)\citenamefont
  {Bauters}, \citenamefont {Heck}, \citenamefont {John}, \citenamefont
  {Barton}, \citenamefont {Bruinink}, \citenamefont {Leinse}, \citenamefont
  {Heideman}, \citenamefont {Blumenthal},\ and\ \citenamefont
  {Bowers}}]{Bauters:11}%
  \BibitemOpen
  \bibfield  {author} {\bibinfo {author} {\bibfnamefont {J.~F.}\ \bibnamefont
  {Bauters}}, \bibinfo {author} {\bibfnamefont {M.~J.~R.}\ \bibnamefont
  {Heck}}, \bibinfo {author} {\bibfnamefont {D.~D.}\ \bibnamefont {John}},
  \bibinfo {author} {\bibfnamefont {J.~S.}\ \bibnamefont {Barton}}, \bibinfo
  {author} {\bibfnamefont {C.~M.}\ \bibnamefont {Bruinink}}, \bibinfo {author}
  {\bibfnamefont {A.}~\bibnamefont {Leinse}}, \bibinfo {author} {\bibfnamefont
  {R.~G.}\ \bibnamefont {Heideman}}, \bibinfo {author} {\bibfnamefont {D.~J.}\
  \bibnamefont {Blumenthal}},\ and\ \bibinfo {author} {\bibfnamefont {J.~E.}\
  \bibnamefont {Bowers}},\ }\bibfield  {title} {\bibinfo {title} {Planar
  waveguides with less than 0.1 {dB}/m propagation loss fabricated with wafer
  bonding},\ }\href {https://doi.org/10.1364/OE.19.024090} {\bibfield
  {journal} {\bibinfo  {journal} {Opt. Express}\ }\textbf {\bibinfo {volume}
  {19}},\ \bibinfo {pages} {24090} (\bibinfo {year} {2011})}\BibitemShut
  {NoStop}%
\bibitem [{\citenamefont {Liu}\ \emph {et~al.}(2021)\citenamefont {Liu},
  \citenamefont {Huang}, \citenamefont {Wang}, \citenamefont {He},
  \citenamefont {Raja}, \citenamefont {Liu}, \citenamefont {Engelsen},\ and\
  \citenamefont {Kippenberg}}]{liu2021high}%
  \BibitemOpen
  \bibfield  {author} {\bibinfo {author} {\bibfnamefont {J.}~\bibnamefont
  {Liu}}, \bibinfo {author} {\bibfnamefont {G.}~\bibnamefont {Huang}}, \bibinfo
  {author} {\bibfnamefont {R.~N.}\ \bibnamefont {Wang}}, \bibinfo {author}
  {\bibfnamefont {J.}~\bibnamefont {He}}, \bibinfo {author} {\bibfnamefont
  {A.~S.}\ \bibnamefont {Raja}}, \bibinfo {author} {\bibfnamefont
  {T.}~\bibnamefont {Liu}}, \bibinfo {author} {\bibfnamefont {N.~J.}\
  \bibnamefont {Engelsen}},\ and\ \bibinfo {author} {\bibfnamefont {T.~J.}\
  \bibnamefont {Kippenberg}},\ }\bibfield  {title} {\bibinfo {title}
  {High-yield, wafer-scale fabrication of ultralow-loss, dispersion-engineered
  silicon nitride photonic circuits},\ }\href
  {https://www.nature.com/articles/s41467-021-21973-z} {\bibfield  {journal}
  {\bibinfo  {journal} {Nature communications}\ }\textbf {\bibinfo {volume}
  {12}},\ \bibinfo {pages} {1} (\bibinfo {year} {2021})}\BibitemShut {NoStop}%
\bibitem [{\citenamefont {Ji}\ \emph {et~al.}(2021)\citenamefont {Ji},
  \citenamefont {Roberts}, \citenamefont {Corato-Zanarella},\ and\
  \citenamefont {Lipson}}]{ji2021methods}%
  \BibitemOpen
  \bibfield  {author} {\bibinfo {author} {\bibfnamefont {X.}~\bibnamefont
  {Ji}}, \bibinfo {author} {\bibfnamefont {S.}~\bibnamefont {Roberts}},
  \bibinfo {author} {\bibfnamefont {M.}~\bibnamefont {Corato-Zanarella}},\ and\
  \bibinfo {author} {\bibfnamefont {M.}~\bibnamefont {Lipson}},\ }\bibfield
  {title} {\bibinfo {title} {Methods to achieve ultra-high quality factor
  silicon nitride resonators},\ }\href
  {https://aip.scitation.org/doi/10.1063/5.0057881} {\bibfield  {journal}
  {\bibinfo  {journal} {APL Photonics}\ }\textbf {\bibinfo {volume} {6}},\
  \bibinfo {pages} {071101} (\bibinfo {year} {2021})}\BibitemShut {NoStop}%
\bibitem [{\citenamefont {Puckett}\ \emph {et~al.}(2021)\citenamefont
  {Puckett}, \citenamefont {Liu}, \citenamefont {Chauhan}, \citenamefont
  {Zhao}, \citenamefont {Jin}, \citenamefont {Cheng}, \citenamefont {Wu},
  \citenamefont {Behunin}, \citenamefont {Rakich}, \citenamefont {Nelson},\
  and\ \citenamefont {Blumenthal}}]{Puckett2021422}%
  \BibitemOpen
  \bibfield  {author} {\bibinfo {author} {\bibfnamefont {M.~W.}\ \bibnamefont
  {Puckett}}, \bibinfo {author} {\bibfnamefont {K.}~\bibnamefont {Liu}},
  \bibinfo {author} {\bibfnamefont {N.}~\bibnamefont {Chauhan}}, \bibinfo
  {author} {\bibfnamefont {Q.}~\bibnamefont {Zhao}}, \bibinfo {author}
  {\bibfnamefont {N.}~\bibnamefont {Jin}}, \bibinfo {author} {\bibfnamefont
  {H.}~\bibnamefont {Cheng}}, \bibinfo {author} {\bibfnamefont
  {J.}~\bibnamefont {Wu}}, \bibinfo {author} {\bibfnamefont {R.~O.}\
  \bibnamefont {Behunin}}, \bibinfo {author} {\bibfnamefont {P.~T.}\
  \bibnamefont {Rakich}}, \bibinfo {author} {\bibfnamefont {K.~D.}\
  \bibnamefont {Nelson}},\ and\ \bibinfo {author} {\bibfnamefont {D.~J.}\
  \bibnamefont {Blumenthal}},\ }\bibfield  {title} {\bibinfo {title} {422
  {Million} intrinsic quality factor planar integrated all-waveguide resonator
  with {sub-MHz} linewidth},\ }\href
  {https://doi.org/10.1038/s41467-021-21205-4} {\bibfield  {journal} {\bibinfo
  {journal} {Nature Communications}\ }\textbf {\bibinfo {volume} {12}},\
  \bibinfo {pages} {934} (\bibinfo {year} {2021})}\BibitemShut {NoStop}%
\bibitem [{\citenamefont {Shulaker}\ \emph {et~al.}(2017)\citenamefont
  {Shulaker}, \citenamefont {Hills}, \citenamefont {Park}, \citenamefont
  {Howe}, \citenamefont {Saraswat}, \citenamefont {Wong},\ and\ \citenamefont
  {Mitra}}]{shulaker2017three}%
  \BibitemOpen
  \bibfield  {author} {\bibinfo {author} {\bibfnamefont {M.~M.}\ \bibnamefont
  {Shulaker}}, \bibinfo {author} {\bibfnamefont {G.}~\bibnamefont {Hills}},
  \bibinfo {author} {\bibfnamefont {R.~S.}\ \bibnamefont {Park}}, \bibinfo
  {author} {\bibfnamefont {R.~T.}\ \bibnamefont {Howe}}, \bibinfo {author}
  {\bibfnamefont {K.}~\bibnamefont {Saraswat}}, \bibinfo {author}
  {\bibfnamefont {H.-S.~P.}\ \bibnamefont {Wong}},\ and\ \bibinfo {author}
  {\bibfnamefont {S.}~\bibnamefont {Mitra}},\ }\bibfield  {title} {\bibinfo
  {title} {Three-dimensional integration of nanotechnologies for computing and
  data storage on a single chip},\ }\href
  {https://www.nature.com/articles/nature22994} {\bibfield  {journal} {\bibinfo
   {journal} {Nature}\ }\textbf {\bibinfo {volume} {547}},\ \bibinfo {pages}
  {74} (\bibinfo {year} {2017})}\BibitemShut {NoStop}%
\bibitem [{\citenamefont {Rachmady}\ \emph {et~al.}(2019)\citenamefont
  {Rachmady}, \citenamefont {Agrawal}, \citenamefont {Sung}, \citenamefont
  {Dewey}, \citenamefont {Chouksey}, \citenamefont {Chu-Kung}, \citenamefont
  {Elbaz}, \citenamefont {Fischer}, \citenamefont {Huang}, \citenamefont {Jun},
  \citenamefont {Krist}, \citenamefont {Metz}, \citenamefont {Michaelos},
  \citenamefont {Mueller}, \citenamefont {Oni}, \citenamefont {Paul},
  \citenamefont {Phan}, \citenamefont {Sears}, \citenamefont {Talukdar},
  \citenamefont {Torres}, \citenamefont {Turkot}, \citenamefont {Wong},
  \citenamefont {Yoo},\ and\ \citenamefont {Kavalieros}}]{Rachmady2019300mm}%
  \BibitemOpen
  \bibfield  {author} {\bibinfo {author} {\bibfnamefont {W.}~\bibnamefont
  {Rachmady}}, \bibinfo {author} {\bibfnamefont {A.}~\bibnamefont {Agrawal}},
  \bibinfo {author} {\bibfnamefont {S.}~\bibnamefont {Sung}}, \bibinfo {author}
  {\bibfnamefont {G.}~\bibnamefont {Dewey}}, \bibinfo {author} {\bibfnamefont
  {S.}~\bibnamefont {Chouksey}}, \bibinfo {author} {\bibfnamefont
  {B.}~\bibnamefont {Chu-Kung}}, \bibinfo {author} {\bibfnamefont
  {G.}~\bibnamefont {Elbaz}}, \bibinfo {author} {\bibfnamefont
  {P.}~\bibnamefont {Fischer}}, \bibinfo {author} {\bibfnamefont {C.~Y.}\
  \bibnamefont {Huang}}, \bibinfo {author} {\bibfnamefont {K.}~\bibnamefont
  {Jun}}, \bibinfo {author} {\bibfnamefont {B.}~\bibnamefont {Krist}}, \bibinfo
  {author} {\bibfnamefont {M.}~\bibnamefont {Metz}}, \bibinfo {author}
  {\bibfnamefont {T.}~\bibnamefont {Michaelos}}, \bibinfo {author}
  {\bibfnamefont {B.}~\bibnamefont {Mueller}}, \bibinfo {author} {\bibfnamefont
  {A.~A.}\ \bibnamefont {Oni}}, \bibinfo {author} {\bibfnamefont
  {R.}~\bibnamefont {Paul}}, \bibinfo {author} {\bibfnamefont {A.}~\bibnamefont
  {Phan}}, \bibinfo {author} {\bibfnamefont {P.}~\bibnamefont {Sears}},
  \bibinfo {author} {\bibfnamefont {T.}~\bibnamefont {Talukdar}}, \bibinfo
  {author} {\bibfnamefont {J.}~\bibnamefont {Torres}}, \bibinfo {author}
  {\bibfnamefont {R.}~\bibnamefont {Turkot}}, \bibinfo {author} {\bibfnamefont
  {L.}~\bibnamefont {Wong}}, \bibinfo {author} {\bibfnamefont {H.~J.}\
  \bibnamefont {Yoo}},\ and\ \bibinfo {author} {\bibfnamefont {J.}~\bibnamefont
  {Kavalieros}},\ }\bibfield  {title} {\bibinfo {title} {300mm heterogeneous 3D
  integration of record performance layer transfer germanium PMOS with silicon
  NMOS for low power high performance logic applications},\ }in\ \href
  {https://ieeexplore.ieee.org/document/8993626} {\emph {\bibinfo {booktitle}
  {2019 IEEE International Electron Devices Meeting (IEDM)}}}\ (\bibinfo
  {organization} {IEEE},\ \bibinfo {year} {2019})\ pp.\ \bibinfo {pages}
  {29--7}\BibitemShut {NoStop}%
\bibitem [{\citenamefont {Sacher}\ \emph {et~al.}(2018)\citenamefont {Sacher},
  \citenamefont {Mikkelsen}, \citenamefont {Huang}, \citenamefont {Mak},
  \citenamefont {Yong}, \citenamefont {Luo}, \citenamefont {Li}, \citenamefont
  {Dumais}, \citenamefont {Jiang}, \citenamefont {Goodwill} \emph
  {et~al.}}]{sacher2018monolithically}%
  \BibitemOpen
  \bibfield  {author} {\bibinfo {author} {\bibfnamefont {W.~D.}\ \bibnamefont
  {Sacher}}, \bibinfo {author} {\bibfnamefont {J.~C.}\ \bibnamefont
  {Mikkelsen}}, \bibinfo {author} {\bibfnamefont {Y.}~\bibnamefont {Huang}},
  \bibinfo {author} {\bibfnamefont {J.~C.}\ \bibnamefont {Mak}}, \bibinfo
  {author} {\bibfnamefont {Z.}~\bibnamefont {Yong}}, \bibinfo {author}
  {\bibfnamefont {X.}~\bibnamefont {Luo}}, \bibinfo {author} {\bibfnamefont
  {Y.}~\bibnamefont {Li}}, \bibinfo {author} {\bibfnamefont {P.}~\bibnamefont
  {Dumais}}, \bibinfo {author} {\bibfnamefont {J.}~\bibnamefont {Jiang}},
  \bibinfo {author} {\bibfnamefont {D.}~\bibnamefont {Goodwill}}, \emph
  {et~al.},\ }\bibfield  {title} {\bibinfo {title} {Monolithically integrated
  multilayer silicon nitride-on-silicon waveguide platforms for 3-{D} photonic
  circuits and devices},\ }\href {https://ieeexplore.ieee.org/document/8452165}
  {\bibfield  {journal} {\bibinfo  {journal} {Proceedings of the IEEE}\
  }\textbf {\bibinfo {volume} {106}},\ \bibinfo {pages} {2232} (\bibinfo {year}
  {2018})}\BibitemShut {NoStop}%
\bibitem [{\citenamefont {{Komljenovic}}\ \emph {et~al.}(2016)\citenamefont
  {{Komljenovic}}, \citenamefont {{Davenport}}, \citenamefont {{Hulme}},
  \citenamefont {{Liu}}, \citenamefont {{Santis}}, \citenamefont {{Spott}},
  \citenamefont {{Srinivasan}}, \citenamefont {{Stanton}}, \citenamefont
  {{Zhang}},\ and\ \citenamefont {{Bowers}}}]{Komljenovic:16}%
  \BibitemOpen
  \bibfield  {author} {\bibinfo {author} {\bibfnamefont {T.}~\bibnamefont
  {{Komljenovic}}}, \bibinfo {author} {\bibfnamefont {M.}~\bibnamefont
  {{Davenport}}}, \bibinfo {author} {\bibfnamefont {J.}~\bibnamefont
  {{Hulme}}}, \bibinfo {author} {\bibfnamefont {A.~Y.}\ \bibnamefont {{Liu}}},
  \bibinfo {author} {\bibfnamefont {C.~T.}\ \bibnamefont {{Santis}}}, \bibinfo
  {author} {\bibfnamefont {A.}~\bibnamefont {{Spott}}}, \bibinfo {author}
  {\bibfnamefont {S.}~\bibnamefont {{Srinivasan}}}, \bibinfo {author}
  {\bibfnamefont {E.~J.}\ \bibnamefont {{Stanton}}}, \bibinfo {author}
  {\bibfnamefont {C.}~\bibnamefont {{Zhang}}},\ and\ \bibinfo {author}
  {\bibfnamefont {J.~E.}\ \bibnamefont {{Bowers}}},\ }\bibfield  {title}
  {\bibinfo {title} {Heterogeneous Silicon Photonic Integrated Circuits},\
  }\href {https://doi.org/10.1109/JLT.2015.2465382} {\bibfield  {journal}
  {\bibinfo  {journal} {Journal of Lightwave Technology}\ }\textbf {\bibinfo
  {volume} {34}},\ \bibinfo {pages} {20} (\bibinfo {year} {2016})}\BibitemShut
  {NoStop}%
\bibitem [{\citenamefont {Xiang}\ \emph {et~al.}(2020)\citenamefont {Xiang},
  \citenamefont {Jin}, \citenamefont {Guo}, \citenamefont {Peters},
  \citenamefont {Kennedy}, \citenamefont {Selvidge}, \citenamefont {Morton},\
  and\ \citenamefont {Bowers}}]{Xiang:20}%
  \BibitemOpen
  \bibfield  {author} {\bibinfo {author} {\bibfnamefont {C.}~\bibnamefont
  {Xiang}}, \bibinfo {author} {\bibfnamefont {W.}~\bibnamefont {Jin}}, \bibinfo
  {author} {\bibfnamefont {J.}~\bibnamefont {Guo}}, \bibinfo {author}
  {\bibfnamefont {J.~D.}\ \bibnamefont {Peters}}, \bibinfo {author}
  {\bibfnamefont {M.~J.}\ \bibnamefont {Kennedy}}, \bibinfo {author}
  {\bibfnamefont {J.}~\bibnamefont {Selvidge}}, \bibinfo {author}
  {\bibfnamefont {P.~A.}\ \bibnamefont {Morton}},\ and\ \bibinfo {author}
  {\bibfnamefont {J.~E.}\ \bibnamefont {Bowers}},\ }\bibfield  {title}
  {\bibinfo {title} {Narrow-linewidth {III-V/Si/Si$_3$N$_4$} laser using
  multilayer heterogeneous integration},\ }\href
  {https://doi.org/10.1364/OPTICA.384026} {\bibfield  {journal} {\bibinfo
  {journal} {Optica}\ }\textbf {\bibinfo {volume} {7}},\ \bibinfo {pages} {20}
  (\bibinfo {year} {2020})}\BibitemShut {NoStop}%
\bibitem [{\citenamefont {Xiang}\ \emph {et~al.}(2021)\citenamefont {Xiang},
  \citenamefont {Liu}, \citenamefont {Guo}, \citenamefont {Chang},
  \citenamefont {Wang}, \citenamefont {Weng}, \citenamefont {Peters},
  \citenamefont {Xie}, \citenamefont {Zhang}, \citenamefont {Riemensberger},
  \citenamefont {Selvidge}, \citenamefont {Kippenberg},\ and\ \citenamefont
  {Bowers}}]{xiang2021laser}%
  \BibitemOpen
  \bibfield  {author} {\bibinfo {author} {\bibfnamefont {C.}~\bibnamefont
  {Xiang}}, \bibinfo {author} {\bibfnamefont {J.}~\bibnamefont {Liu}}, \bibinfo
  {author} {\bibfnamefont {J.}~\bibnamefont {Guo}}, \bibinfo {author}
  {\bibfnamefont {L.}~\bibnamefont {Chang}}, \bibinfo {author} {\bibfnamefont
  {R.~N.}\ \bibnamefont {Wang}}, \bibinfo {author} {\bibfnamefont
  {W.}~\bibnamefont {Weng}}, \bibinfo {author} {\bibfnamefont {J.}~\bibnamefont
  {Peters}}, \bibinfo {author} {\bibfnamefont {W.}~\bibnamefont {Xie}},
  \bibinfo {author} {\bibfnamefont {Z.}~\bibnamefont {Zhang}}, \bibinfo
  {author} {\bibfnamefont {J.}~\bibnamefont {Riemensberger}}, \bibinfo {author}
  {\bibfnamefont {J.}~\bibnamefont {Selvidge}}, \bibinfo {author}
  {\bibfnamefont {T.~J.}\ \bibnamefont {Kippenberg}},\ and\ \bibinfo {author}
  {\bibfnamefont {J.~E.}\ \bibnamefont {Bowers}},\ }\bibfield  {title}
  {\bibinfo {title} {Laser soliton microcombs heterogeneously integrated on
  silicon},\ }\href {https://www.science.org/doi/abs/10.1126/science.abh2076}
  {\bibfield  {journal} {\bibinfo  {journal} {Science}\ }\textbf {\bibinfo
  {volume} {373}},\ \bibinfo {pages} {99} (\bibinfo {year} {2021})}\BibitemShut
  {NoStop}%
\bibitem [{\citenamefont {Kondratiev}\ \emph {et~al.}(2022)\citenamefont
  {Kondratiev}, \citenamefont {Lobanov}, \citenamefont {Shitikov},
  \citenamefont {Galiev}, \citenamefont {Chermoshentsev}, \citenamefont
  {Dmitriev}, \citenamefont {Danilin}, \citenamefont {Lonshakov}, \citenamefont
  {Min'kov}, \citenamefont {Sokol} \emph {et~al.}}]{kondratiev2022recent}%
  \BibitemOpen
  \bibfield  {author} {\bibinfo {author} {\bibfnamefont {N.~M.}\ \bibnamefont
  {Kondratiev}}, \bibinfo {author} {\bibfnamefont {V.~E.}\ \bibnamefont
  {Lobanov}}, \bibinfo {author} {\bibfnamefont {A.~E.}\ \bibnamefont
  {Shitikov}}, \bibinfo {author} {\bibfnamefont {R.~R.}\ \bibnamefont
  {Galiev}}, \bibinfo {author} {\bibfnamefont {D.~A.}\ \bibnamefont
  {Chermoshentsev}}, \bibinfo {author} {\bibfnamefont {N.~Y.}\ \bibnamefont
  {Dmitriev}}, \bibinfo {author} {\bibfnamefont {A.~N.}\ \bibnamefont
  {Danilin}}, \bibinfo {author} {\bibfnamefont {E.~A.}\ \bibnamefont
  {Lonshakov}}, \bibinfo {author} {\bibfnamefont {K.~N.}\ \bibnamefont
  {Min'kov}}, \bibinfo {author} {\bibfnamefont {D.~M.}\ \bibnamefont {Sokol}},
  \emph {et~al.},\ }\bibfield  {title} {\bibinfo {title} {Recent Advances in
  Laser Self-Injection Locking to High-$ Q $ Microresonators},\ }\href
  {https://arxiv.org/abs/2212.05730} {\bibfield  {journal} {\bibinfo  {journal}
  {arXiv preprint arXiv:2212.05730}\ } (\bibinfo {year} {2022})}\BibitemShut
  {NoStop}%
\bibitem [{\citenamefont {Liang}\ \emph {et~al.}(2015)\citenamefont {Liang},
  \citenamefont {Ilchenko}, \citenamefont {Eliyahu}, \citenamefont
  {Savchenkov}, \citenamefont {Matsko}, \citenamefont {Seidel},\ and\
  \citenamefont {Maleki}}]{Liang:15b}%
  \BibitemOpen
  \bibfield  {author} {\bibinfo {author} {\bibfnamefont {W.}~\bibnamefont
  {Liang}}, \bibinfo {author} {\bibfnamefont {V.~S.}\ \bibnamefont {Ilchenko}},
  \bibinfo {author} {\bibfnamefont {D.}~\bibnamefont {Eliyahu}}, \bibinfo
  {author} {\bibfnamefont {A.~A.}\ \bibnamefont {Savchenkov}}, \bibinfo
  {author} {\bibfnamefont {A.~B.}\ \bibnamefont {Matsko}}, \bibinfo {author}
  {\bibfnamefont {D.}~\bibnamefont {Seidel}},\ and\ \bibinfo {author}
  {\bibfnamefont {L.}~\bibnamefont {Maleki}},\ }\bibfield  {title} {\bibinfo
  {title} {Ultralow noise miniature external cavity semiconductor laser},\
  }\href {https://doi.org/10.1038/ncomms8371} {\bibfield  {journal} {\bibinfo
  {journal} {Nature Communications}\ }\textbf {\bibinfo {volume} {6}},\
  \bibinfo {pages} {7371} (\bibinfo {year} {2015})}\BibitemShut {NoStop}%
\bibitem [{\citenamefont {Jin}\ \emph {et~al.}(2021)\citenamefont {Jin},
  \citenamefont {Yang}, \citenamefont {Chang}, \citenamefont {Shen},
  \citenamefont {Wang}, \citenamefont {Leal}, \citenamefont {Wu}, \citenamefont
  {Gao}, \citenamefont {Feshali}, \citenamefont {Paniccia}, \citenamefont
  {Vahala},\ and\ \citenamefont {Bowers}}]{Jin2021hertz}%
  \BibitemOpen
  \bibfield  {author} {\bibinfo {author} {\bibfnamefont {W.}~\bibnamefont
  {Jin}}, \bibinfo {author} {\bibfnamefont {Q.-F.}\ \bibnamefont {Yang}},
  \bibinfo {author} {\bibfnamefont {L.}~\bibnamefont {Chang}}, \bibinfo
  {author} {\bibfnamefont {B.}~\bibnamefont {Shen}}, \bibinfo {author}
  {\bibfnamefont {H.}~\bibnamefont {Wang}}, \bibinfo {author} {\bibfnamefont
  {M.~A.}\ \bibnamefont {Leal}}, \bibinfo {author} {\bibfnamefont
  {L.}~\bibnamefont {Wu}}, \bibinfo {author} {\bibfnamefont {M.}~\bibnamefont
  {Gao}}, \bibinfo {author} {\bibfnamefont {A.}~\bibnamefont {Feshali}},
  \bibinfo {author} {\bibfnamefont {M.}~\bibnamefont {Paniccia}}, \bibinfo
  {author} {\bibfnamefont {K.~J.}\ \bibnamefont {Vahala}},\ and\ \bibinfo
  {author} {\bibfnamefont {J.~E.}\ \bibnamefont {Bowers}},\ }\bibfield  {title}
  {\bibinfo {title} {Hertz-linewidth semiconductor lasers using CMOS-ready
  ultra-high-Q microresonators},\ }\href
  {https://doi.org/10.1038/s41566-021-00761-7} {\bibfield  {journal} {\bibinfo
  {journal} {Nature Photonics}\ }\textbf {\bibinfo {volume} {15}},\ \bibinfo
  {pages} {346} (\bibinfo {year} {2021})}\BibitemShut {NoStop}%
\bibitem [{\citenamefont {Li}\ \emph {et~al.}(2021)\citenamefont {Li},
  \citenamefont {Jin}, \citenamefont {Wu}, \citenamefont {Chang}, \citenamefont
  {Wang}, \citenamefont {Shen}, \citenamefont {Yuan}, \citenamefont {Feshali},
  \citenamefont {Paniccia}, \citenamefont {Vahala},\ and\ \citenamefont
  {Bowers}}]{li2021reaching}%
  \BibitemOpen
  \bibfield  {author} {\bibinfo {author} {\bibfnamefont {B.}~\bibnamefont
  {Li}}, \bibinfo {author} {\bibfnamefont {W.}~\bibnamefont {Jin}}, \bibinfo
  {author} {\bibfnamefont {L.}~\bibnamefont {Wu}}, \bibinfo {author}
  {\bibfnamefont {L.}~\bibnamefont {Chang}}, \bibinfo {author} {\bibfnamefont
  {H.}~\bibnamefont {Wang}}, \bibinfo {author} {\bibfnamefont {B.}~\bibnamefont
  {Shen}}, \bibinfo {author} {\bibfnamefont {Z.}~\bibnamefont {Yuan}}, \bibinfo
  {author} {\bibfnamefont {A.}~\bibnamefont {Feshali}}, \bibinfo {author}
  {\bibfnamefont {M.}~\bibnamefont {Paniccia}}, \bibinfo {author}
  {\bibfnamefont {K.~J.}\ \bibnamefont {Vahala}},\ and\ \bibinfo {author}
  {\bibfnamefont {J.~E.}\ \bibnamefont {Bowers}},\ }\bibfield  {title}
  {\bibinfo {title} {Reaching fiber-laser coherence in integrated photonics},\
  }\href {https://opg.optica.org/ol/abstract.cfm?uri=ol-46-20-5201} {\bibfield
  {journal} {\bibinfo  {journal} {Optics Letters}\ }\textbf {\bibinfo {volume}
  {46}},\ \bibinfo {pages} {5201} (\bibinfo {year} {2021})}\BibitemShut
  {NoStop}%
\bibitem [{\citenamefont {Leeson}\ and\ \citenamefont
  {Johnson}(1966)}]{leeson1966short}%
  \BibitemOpen
  \bibfield  {author} {\bibinfo {author} {\bibfnamefont {D.}~\bibnamefont
  {Leeson}}\ and\ \bibinfo {author} {\bibfnamefont {G.}~\bibnamefont
  {Johnson}},\ }\bibfield  {title} {\bibinfo {title} {Short-term stability for
  a Doppler radar: Requirements, measurements, and techniques},\ }\href
  {https://ieeexplore.ieee.org/abstract/document/1446567} {\bibfield  {journal}
  {\bibinfo  {journal} {Proceedings of the IEEE}\ }\textbf {\bibinfo {volume}
  {54}},\ \bibinfo {pages} {244} (\bibinfo {year} {1966})}\BibitemShut
  {NoStop}%
\bibitem [{\citenamefont {Schunk}\ and\ \citenamefont
  {Petermann}(1988)}]{schunk1988numerical}%
  \BibitemOpen
  \bibfield  {author} {\bibinfo {author} {\bibfnamefont {N.}~\bibnamefont
  {Schunk}}\ and\ \bibinfo {author} {\bibfnamefont {K.}~\bibnamefont
  {Petermann}},\ }\bibfield  {title} {\bibinfo {title} {Numerical analysis of
  the feedback regimes for a single-mode semiconductor laser with external
  feedback},\ }\href {https://ieeexplore.ieee.org/document/960} {\bibfield
  {journal} {\bibinfo  {journal} {IEEE Journal of Quantum Electronics}\
  }\textbf {\bibinfo {volume} {24}},\ \bibinfo {pages} {1242} (\bibinfo {year}
  {1988})}\BibitemShut {NoStop}%
\bibitem [{\citenamefont {Zhang}\ \emph {et~al.}(2020)\citenamefont {Zhang},
  \citenamefont {Zou}, \citenamefont {Wang}, \citenamefont {Liao},
  \citenamefont {Satyan}, \citenamefont {Rakuljic}, \citenamefont {Willner},\
  and\ \citenamefont {Yariv}}]{zhang2020high}%
  \BibitemOpen
  \bibfield  {author} {\bibinfo {author} {\bibfnamefont {Z.}~\bibnamefont
  {Zhang}}, \bibinfo {author} {\bibfnamefont {K.}~\bibnamefont {Zou}}, \bibinfo
  {author} {\bibfnamefont {H.}~\bibnamefont {Wang}}, \bibinfo {author}
  {\bibfnamefont {P.}~\bibnamefont {Liao}}, \bibinfo {author} {\bibfnamefont
  {N.}~\bibnamefont {Satyan}}, \bibinfo {author} {\bibfnamefont
  {G.}~\bibnamefont {Rakuljic}}, \bibinfo {author} {\bibfnamefont {A.~E.}\
  \bibnamefont {Willner}},\ and\ \bibinfo {author} {\bibfnamefont
  {A.}~\bibnamefont {Yariv}},\ }\bibfield  {title} {\bibinfo {title}
  {High-speed coherent optical communication with isolator-free heterogeneous
  Si/III-V lasers},\ }\href
  {https://ieeexplore.ieee.org/abstract/document/9165107} {\bibfield  {journal}
  {\bibinfo  {journal} {Journal of Lightwave Technology}\ }\textbf {\bibinfo
  {volume} {38}},\ \bibinfo {pages} {6584} (\bibinfo {year}
  {2020})}\BibitemShut {NoStop}%
\bibitem [{\citenamefont {Gomez}\ \emph {et~al.}(2020)\citenamefont {Gomez},
  \citenamefont {Huang}, \citenamefont {Duan}, \citenamefont {Combri{\'e}},
  \citenamefont {Shen}, \citenamefont {Baili}, \citenamefont {de~Rossi},\ and\
  \citenamefont {Grillot}}]{gomez2020high}%
  \BibitemOpen
  \bibfield  {author} {\bibinfo {author} {\bibfnamefont {S.}~\bibnamefont
  {Gomez}}, \bibinfo {author} {\bibfnamefont {H.}~\bibnamefont {Huang}},
  \bibinfo {author} {\bibfnamefont {J.}~\bibnamefont {Duan}}, \bibinfo {author}
  {\bibfnamefont {S.}~\bibnamefont {Combri{\'e}}}, \bibinfo {author}
  {\bibfnamefont {A.}~\bibnamefont {Shen}}, \bibinfo {author} {\bibfnamefont
  {G.}~\bibnamefont {Baili}}, \bibinfo {author} {\bibfnamefont
  {A.}~\bibnamefont {de~Rossi}},\ and\ \bibinfo {author} {\bibfnamefont
  {F.}~\bibnamefont {Grillot}},\ }\bibfield  {title} {\bibinfo {title} {High
  coherence collapse of a hybrid III--V/Si semiconductor laser with a large
  quality factor},\ }\href
  {https://iopscience.iop.org/article/10.1088/2515-7647/ab6a74} {\bibfield
  {journal} {\bibinfo  {journal} {Journal of Physics: Photonics}\ }\textbf
  {\bibinfo {volume} {2}},\ \bibinfo {pages} {025005} (\bibinfo {year}
  {2020})}\BibitemShut {NoStop}%
\bibitem [{\citenamefont {Guo}\ \emph {et~al.}(2022)\citenamefont {Guo},
  \citenamefont {McLemore}, \citenamefont {Xiang}, \citenamefont {Lee},
  \citenamefont {Wu}, \citenamefont {Jin}, \citenamefont {Kelleher},
  \citenamefont {Jin}, \citenamefont {Mason}, \citenamefont {Chang},
  \citenamefont {Feshali}, \citenamefont {Paniccia}, \citenamefont {Rakich},
  \citenamefont {Vahala}, \citenamefont {Diddams}, \citenamefont {Quinlan},\
  and\ \citenamefont {Bowers}}]{guo2022chip}%
  \BibitemOpen
  \bibfield  {author} {\bibinfo {author} {\bibfnamefont {J.}~\bibnamefont
  {Guo}}, \bibinfo {author} {\bibfnamefont {C.~A.}\ \bibnamefont {McLemore}},
  \bibinfo {author} {\bibfnamefont {C.}~\bibnamefont {Xiang}}, \bibinfo
  {author} {\bibfnamefont {D.}~\bibnamefont {Lee}}, \bibinfo {author}
  {\bibfnamefont {L.}~\bibnamefont {Wu}}, \bibinfo {author} {\bibfnamefont
  {W.}~\bibnamefont {Jin}}, \bibinfo {author} {\bibfnamefont {M.}~\bibnamefont
  {Kelleher}}, \bibinfo {author} {\bibfnamefont {N.}~\bibnamefont {Jin}},
  \bibinfo {author} {\bibfnamefont {D.}~\bibnamefont {Mason}}, \bibinfo
  {author} {\bibfnamefont {L.}~\bibnamefont {Chang}}, \bibinfo {author}
  {\bibfnamefont {A.}~\bibnamefont {Feshali}}, \bibinfo {author} {\bibfnamefont
  {M.}~\bibnamefont {Paniccia}}, \bibinfo {author} {\bibfnamefont {P.~T.}\
  \bibnamefont {Rakich}}, \bibinfo {author} {\bibfnamefont {K.~J.}\
  \bibnamefont {Vahala}}, \bibinfo {author} {\bibfnamefont {S.~A.}\
  \bibnamefont {Diddams}}, \bibinfo {author} {\bibfnamefont {F.}~\bibnamefont
  {Quinlan}},\ and\ \bibinfo {author} {\bibfnamefont {J.~E.}\ \bibnamefont
  {Bowers}},\ }\bibfield  {title} {\bibinfo {title} {Chip-based laser with
  1-hertz integrated linewidth},\ }\href
  {https://www.science.org/doi/abs/10.1126/sciadv.abp9006} {\bibfield
  {journal} {\bibinfo  {journal} {Science Advances}\ }\textbf {\bibinfo
  {volume} {8}},\ \bibinfo {pages} {eabp9006} (\bibinfo {year}
  {2022})}\BibitemShut {NoStop}%
\bibitem [{\citenamefont {McLemore}\ \emph {et~al.}(2022)\citenamefont
  {McLemore}, \citenamefont {Jin}, \citenamefont {Kelleher}, \citenamefont
  {Hendrie}, \citenamefont {Mason}, \citenamefont {Luo}, \citenamefont {Lee},
  \citenamefont {Rakich}, \citenamefont {Diddams},\ and\ \citenamefont
  {Quinlan}}]{McLemore2022Miniaturizing}%
  \BibitemOpen
  \bibfield  {author} {\bibinfo {author} {\bibfnamefont {C.~A.}\ \bibnamefont
  {McLemore}}, \bibinfo {author} {\bibfnamefont {N.}~\bibnamefont {Jin}},
  \bibinfo {author} {\bibfnamefont {M.~L.}\ \bibnamefont {Kelleher}}, \bibinfo
  {author} {\bibfnamefont {J.~P.}\ \bibnamefont {Hendrie}}, \bibinfo {author}
  {\bibfnamefont {D.}~\bibnamefont {Mason}}, \bibinfo {author} {\bibfnamefont
  {Y.}~\bibnamefont {Luo}}, \bibinfo {author} {\bibfnamefont {D.}~\bibnamefont
  {Lee}}, \bibinfo {author} {\bibfnamefont {P.}~\bibnamefont {Rakich}},
  \bibinfo {author} {\bibfnamefont {S.~A.}\ \bibnamefont {Diddams}},\ and\
  \bibinfo {author} {\bibfnamefont {F.}~\bibnamefont {Quinlan}},\ }\bibfield
  {title} {\bibinfo {title} {Miniaturizing Ultrastable Electromagnetic
  Oscillators: Sub-${10}^{\ensuremath{-}14}$ Frequency Instability from a
  Centimeter-Scale Fabry-Perot Cavity},\ }\href
  {https://doi.org/10.1103/PhysRevApplied.18.054054} {\bibfield  {journal}
  {\bibinfo  {journal} {Phys. Rev. Applied}\ }\textbf {\bibinfo {volume}
  {18}},\ \bibinfo {pages} {054054} (\bibinfo {year} {2022})}\BibitemShut
  {NoStop}%
\bibitem [{\citenamefont {Jin}\ \emph {et~al.}(2022)\citenamefont {Jin},
  \citenamefont {McLemore}, \citenamefont {Mason}, \citenamefont {Hendrie},
  \citenamefont {Luo}, \citenamefont {Kelleher}, \citenamefont {Kharel},
  \citenamefont {Quinlan}, \citenamefont {Diddams},\ and\ \citenamefont
  {Rakich}}]{Jin2022Micro}%
  \BibitemOpen
  \bibfield  {author} {\bibinfo {author} {\bibfnamefont {N.}~\bibnamefont
  {Jin}}, \bibinfo {author} {\bibfnamefont {C.~A.}\ \bibnamefont {McLemore}},
  \bibinfo {author} {\bibfnamefont {D.}~\bibnamefont {Mason}}, \bibinfo
  {author} {\bibfnamefont {J.~P.}\ \bibnamefont {Hendrie}}, \bibinfo {author}
  {\bibfnamefont {Y.}~\bibnamefont {Luo}}, \bibinfo {author} {\bibfnamefont
  {M.~L.}\ \bibnamefont {Kelleher}}, \bibinfo {author} {\bibfnamefont
  {P.}~\bibnamefont {Kharel}}, \bibinfo {author} {\bibfnamefont
  {F.}~\bibnamefont {Quinlan}}, \bibinfo {author} {\bibfnamefont {S.~A.}\
  \bibnamefont {Diddams}},\ and\ \bibinfo {author} {\bibfnamefont {P.~T.}\
  \bibnamefont {Rakich}},\ }\bibfield  {title} {\bibinfo {title}
  {Micro-fabricated mirrors with finesse exceeding one million},\ }\href
  {https://doi.org/10.1364/OPTICA.467440} {\bibfield  {journal} {\bibinfo
  {journal} {Optica}\ }\textbf {\bibinfo {volume} {9}},\ \bibinfo {pages} {965}
  (\bibinfo {year} {2022})}\BibitemShut {NoStop}%
\bibitem [{\citenamefont {Hulme}\ \emph {et~al.}(2017)\citenamefont {Hulme},
  \citenamefont {Kennedy}, \citenamefont {Chao}, \citenamefont {Liang},
  \citenamefont {Komljenovic}, \citenamefont {Shi}, \citenamefont {Szafraniec},
  \citenamefont {Baney},\ and\ \citenamefont {Bowers}}]{Hulme2017fully}%
  \BibitemOpen
  \bibfield  {author} {\bibinfo {author} {\bibfnamefont {J.}~\bibnamefont
  {Hulme}}, \bibinfo {author} {\bibfnamefont {M.}~\bibnamefont {Kennedy}},
  \bibinfo {author} {\bibfnamefont {R.-L.}\ \bibnamefont {Chao}}, \bibinfo
  {author} {\bibfnamefont {L.}~\bibnamefont {Liang}}, \bibinfo {author}
  {\bibfnamefont {T.}~\bibnamefont {Komljenovic}}, \bibinfo {author}
  {\bibfnamefont {J.-W.}\ \bibnamefont {Shi}}, \bibinfo {author} {\bibfnamefont
  {B.}~\bibnamefont {Szafraniec}}, \bibinfo {author} {\bibfnamefont
  {D.}~\bibnamefont {Baney}},\ and\ \bibinfo {author} {\bibfnamefont {J.~E.}\
  \bibnamefont {Bowers}},\ }\bibfield  {title} {\bibinfo {title} {Fully
  integrated microwave frequency synthesizer on heterogeneous silicon-III/V},\
  }\href {https://doi.org/10.1364/OE.25.002422} {\bibfield  {journal} {\bibinfo
   {journal} {Opt. Express}\ }\textbf {\bibinfo {volume} {25}},\ \bibinfo
  {pages} {2422} (\bibinfo {year} {2017})}\BibitemShut {NoStop}%
\bibitem [{\citenamefont {Kittlaus}\ \emph {et~al.}(2021)\citenamefont
  {Kittlaus}, \citenamefont {Eliyahu}, \citenamefont {Ganji}, \citenamefont
  {Williams}, \citenamefont {Matsko}, \citenamefont {Cooper},\ and\
  \citenamefont {Forouhar}}]{kittlaus2021low}%
  \BibitemOpen
  \bibfield  {author} {\bibinfo {author} {\bibfnamefont {E.~A.}\ \bibnamefont
  {Kittlaus}}, \bibinfo {author} {\bibfnamefont {D.}~\bibnamefont {Eliyahu}},
  \bibinfo {author} {\bibfnamefont {S.}~\bibnamefont {Ganji}}, \bibinfo
  {author} {\bibfnamefont {S.}~\bibnamefont {Williams}}, \bibinfo {author}
  {\bibfnamefont {A.~B.}\ \bibnamefont {Matsko}}, \bibinfo {author}
  {\bibfnamefont {K.~B.}\ \bibnamefont {Cooper}},\ and\ \bibinfo {author}
  {\bibfnamefont {S.}~\bibnamefont {Forouhar}},\ }\bibfield  {title} {\bibinfo
  {title} {A low-noise photonic heterodyne synthesizer and its application to
  millimeter-wave radar},\ }\href
  {https://www.nature.com/articles/s41467-021-24637-0} {\bibfield  {journal}
  {\bibinfo  {journal} {Nature Communications}\ }\textbf {\bibinfo {volume}
  {12}},\ \bibinfo {pages} {1} (\bibinfo {year} {2021})}\BibitemShut {NoStop}%
\bibitem [{\citenamefont {Davenport}\ \emph {et~al.}(2016)\citenamefont
  {Davenport}, \citenamefont {Skend{\v{z}}i{\'c}}, \citenamefont {Volet},
  \citenamefont {Hulme}, \citenamefont {Heck},\ and\ \citenamefont
  {Bowers}}]{davenport2016heterogeneous}%
  \BibitemOpen
  \bibfield  {author} {\bibinfo {author} {\bibfnamefont {M.~L.}\ \bibnamefont
  {Davenport}}, \bibinfo {author} {\bibfnamefont {S.}~\bibnamefont
  {Skend{\v{z}}i{\'c}}}, \bibinfo {author} {\bibfnamefont {N.}~\bibnamefont
  {Volet}}, \bibinfo {author} {\bibfnamefont {J.~C.}\ \bibnamefont {Hulme}},
  \bibinfo {author} {\bibfnamefont {M.~J.}\ \bibnamefont {Heck}},\ and\
  \bibinfo {author} {\bibfnamefont {J.~E.}\ \bibnamefont {Bowers}},\ }\bibfield
   {title} {\bibinfo {title} {Heterogeneous silicon/{III--V} semiconductor
  optical amplifiers},\ }\href {https://ieeexplore.ieee.org/document/7516667}
  {\bibfield  {journal} {\bibinfo  {journal} {IEEE Journal of Selected Topics
  in Quantum Electronics}\ }\textbf {\bibinfo {volume} {22}},\ \bibinfo {pages}
  {78} (\bibinfo {year} {2016})}\BibitemShut {NoStop}%
\bibitem [{\citenamefont {Campbell}(2022)}]{Campbell2022evolution}%
  \BibitemOpen
  \bibfield  {author} {\bibinfo {author} {\bibfnamefont {J.~C.}\ \bibnamefont
  {Campbell}},\ }\bibfield  {title} {\bibinfo {title} {Evolution of Low-Noise
  Avalanche Photodetectors},\ }\href
  {https://doi.org/10.1109/JSTQE.2021.3092963} {\bibfield  {journal} {\bibinfo
  {journal} {IEEE Journal of Selected Topics in Quantum Electronics}\ }\textbf
  {\bibinfo {volume} {28}},\ \bibinfo {pages} {1} (\bibinfo {year}
  {2022})}\BibitemShut {NoStop}%
\bibitem [{\citenamefont {Li}\ \emph {et~al.}(2014)\citenamefont {Li},
  \citenamefont {Yi}, \citenamefont {Lee}, \citenamefont {Diddams},\ and\
  \citenamefont {Vahala}}]{Li:14}%
  \BibitemOpen
  \bibfield  {author} {\bibinfo {author} {\bibfnamefont {J.}~\bibnamefont
  {Li}}, \bibinfo {author} {\bibfnamefont {X.}~\bibnamefont {Yi}}, \bibinfo
  {author} {\bibfnamefont {H.}~\bibnamefont {Lee}}, \bibinfo {author}
  {\bibfnamefont {S.~A.}\ \bibnamefont {Diddams}},\ and\ \bibinfo {author}
  {\bibfnamefont {K.~J.}\ \bibnamefont {Vahala}},\ }\bibfield  {title}
  {\bibinfo {title} {Electro-optical frequency division and stable microwave
  synthesis},\ }\href {https://doi.org/10.1126/science.1252909} {\bibfield
  {journal} {\bibinfo  {journal} {Science}\ }\textbf {\bibinfo {volume}
  {345}},\ \bibinfo {pages} {309} (\bibinfo {year} {2014})}\BibitemShut
  {NoStop}%
\bibitem [{\citenamefont {Gundavarapu}\ \emph {et~al.}(2019)\citenamefont
  {Gundavarapu}, \citenamefont {Brodnik}, \citenamefont {Puckett},
  \citenamefont {Huffman}, \citenamefont {Bose}, \citenamefont {Behunin},
  \citenamefont {Wu}, \citenamefont {Qiu}, \citenamefont {Pinho}, \citenamefont
  {Chauhan}, \citenamefont {Nohava}, \citenamefont {Rakich}, \citenamefont
  {Nelson}, \citenamefont {Salit},\ and\ \citenamefont
  {Blumenthal}}]{Gundavarapu:19}%
  \BibitemOpen
  \bibfield  {author} {\bibinfo {author} {\bibfnamefont {S.}~\bibnamefont
  {Gundavarapu}}, \bibinfo {author} {\bibfnamefont {G.~M.}\ \bibnamefont
  {Brodnik}}, \bibinfo {author} {\bibfnamefont {M.}~\bibnamefont {Puckett}},
  \bibinfo {author} {\bibfnamefont {T.}~\bibnamefont {Huffman}}, \bibinfo
  {author} {\bibfnamefont {D.}~\bibnamefont {Bose}}, \bibinfo {author}
  {\bibfnamefont {R.}~\bibnamefont {Behunin}}, \bibinfo {author} {\bibfnamefont
  {J.}~\bibnamefont {Wu}}, \bibinfo {author} {\bibfnamefont {T.}~\bibnamefont
  {Qiu}}, \bibinfo {author} {\bibfnamefont {C.}~\bibnamefont {Pinho}}, \bibinfo
  {author} {\bibfnamefont {N.}~\bibnamefont {Chauhan}}, \bibinfo {author}
  {\bibfnamefont {J.}~\bibnamefont {Nohava}}, \bibinfo {author} {\bibfnamefont
  {P.~T.}\ \bibnamefont {Rakich}}, \bibinfo {author} {\bibfnamefont {K.~D.}\
  \bibnamefont {Nelson}}, \bibinfo {author} {\bibfnamefont {M.}~\bibnamefont
  {Salit}},\ and\ \bibinfo {author} {\bibfnamefont {D.~J.}\ \bibnamefont
  {Blumenthal}},\ }\bibfield  {title} {\bibinfo {title} {Sub-hertz fundamental
  linewidth photonic integrated {B}rillouin laser},\ }\href
  {https://doi.org/10.1038/s41566-018-0313-2} {\bibfield  {journal} {\bibinfo
  {journal} {Nature Photonics}\ }\textbf {\bibinfo {volume} {13}},\ \bibinfo
  {pages} {60} (\bibinfo {year} {2019})}\BibitemShut {NoStop}%
\bibitem [{\citenamefont {Liu}\ \emph {et~al.}(2022)\citenamefont {Liu},
  \citenamefont {Qiu}, \citenamefont {Ji}, \citenamefont {Lukashchuk},
  \citenamefont {He}, \citenamefont {Riemensberger}, \citenamefont {Hafermann},
  \citenamefont {Wang}, \citenamefont {Liu}, \citenamefont {Ronning},\ and\
  \citenamefont {Kippenberg}}]{liu2022a}%
  \BibitemOpen
  \bibfield  {author} {\bibinfo {author} {\bibfnamefont {Y.}~\bibnamefont
  {Liu}}, \bibinfo {author} {\bibfnamefont {Z.}~\bibnamefont {Qiu}}, \bibinfo
  {author} {\bibfnamefont {X.}~\bibnamefont {Ji}}, \bibinfo {author}
  {\bibfnamefont {A.}~\bibnamefont {Lukashchuk}}, \bibinfo {author}
  {\bibfnamefont {J.}~\bibnamefont {He}}, \bibinfo {author} {\bibfnamefont
  {J.}~\bibnamefont {Riemensberger}}, \bibinfo {author} {\bibfnamefont
  {M.}~\bibnamefont {Hafermann}}, \bibinfo {author} {\bibfnamefont {R.~N.}\
  \bibnamefont {Wang}}, \bibinfo {author} {\bibfnamefont {J.}~\bibnamefont
  {Liu}}, \bibinfo {author} {\bibfnamefont {C.}~\bibnamefont {Ronning}},\ and\
  \bibinfo {author} {\bibfnamefont {T.~J.}\ \bibnamefont {Kippenberg}},\
  }\bibfield  {title} {\bibinfo {title} {A photonic integrated circuit-based
  erbium-doped amplifier},\ }\href
  {https://www.science.org/doi/abs/10.1126/science.abo2631} {\bibfield
  {journal} {\bibinfo  {journal} {Science}\ }\textbf {\bibinfo {volume}
  {376}},\ \bibinfo {pages} {1309} (\bibinfo {year} {2022})}\BibitemShut
  {NoStop}%
\bibitem [{\citenamefont {Liang}\ \emph {et~al.}(2017)\citenamefont {Liang},
  \citenamefont {Ilchenko}, \citenamefont {Savchenkov}, \citenamefont {Dale},
  \citenamefont {Eliyahu}, \citenamefont {Matsko},\ and\ \citenamefont
  {Maleki}}]{Liang:17}%
  \BibitemOpen
  \bibfield  {author} {\bibinfo {author} {\bibfnamefont {W.}~\bibnamefont
  {Liang}}, \bibinfo {author} {\bibfnamefont {V.~S.}\ \bibnamefont {Ilchenko}},
  \bibinfo {author} {\bibfnamefont {A.~A.}\ \bibnamefont {Savchenkov}},
  \bibinfo {author} {\bibfnamefont {E.}~\bibnamefont {Dale}}, \bibinfo {author}
  {\bibfnamefont {D.}~\bibnamefont {Eliyahu}}, \bibinfo {author} {\bibfnamefont
  {A.~B.}\ \bibnamefont {Matsko}},\ and\ \bibinfo {author} {\bibfnamefont
  {L.}~\bibnamefont {Maleki}},\ }\bibfield  {title} {\bibinfo {title} {Resonant
  microphotonic gyroscope},\ }\href {https://doi.org/10.1364/OPTICA.4.000114}
  {\bibfield  {journal} {\bibinfo  {journal} {Optica}\ }\textbf {\bibinfo
  {volume} {4}},\ \bibinfo {pages} {114} (\bibinfo {year} {2017})}\BibitemShut
  {NoStop}%
\bibitem [{\citenamefont {Spencer}\ \emph {et~al.}(2018)\citenamefont
  {Spencer}, \citenamefont {Drake}, \citenamefont {Briles}, \citenamefont
  {Stone}, \citenamefont {Sinclair}, \citenamefont {Fredrick}, \citenamefont
  {Li}, \citenamefont {Westly}, \citenamefont {Ilic}, \citenamefont
  {Bluestone}, \citenamefont {Volet}, \citenamefont {Komljenovic},
  \citenamefont {Chang}, \citenamefont {Lee}, \citenamefont {Oh}, \citenamefont
  {Suh}, \citenamefont {Yang}, \citenamefont {Pfeiffer}, \citenamefont
  {Kippenberg}, \citenamefont {Norberg}, \citenamefont {Theogarajan},
  \citenamefont {Vahala}, \citenamefont {Newbury}, \citenamefont {Srinivasan},
  \citenamefont {Bowers}, \citenamefont {Diddams},\ and\ \citenamefont
  {Papp}}]{Spencer:18}%
  \BibitemOpen
  \bibfield  {author} {\bibinfo {author} {\bibfnamefont {D.~T.}\ \bibnamefont
  {Spencer}}, \bibinfo {author} {\bibfnamefont {T.}~\bibnamefont {Drake}},
  \bibinfo {author} {\bibfnamefont {T.~C.}\ \bibnamefont {Briles}}, \bibinfo
  {author} {\bibfnamefont {J.}~\bibnamefont {Stone}}, \bibinfo {author}
  {\bibfnamefont {L.~C.}\ \bibnamefont {Sinclair}}, \bibinfo {author}
  {\bibfnamefont {C.}~\bibnamefont {Fredrick}}, \bibinfo {author}
  {\bibfnamefont {Q.}~\bibnamefont {Li}}, \bibinfo {author} {\bibfnamefont
  {D.}~\bibnamefont {Westly}}, \bibinfo {author} {\bibfnamefont {B.~R.}\
  \bibnamefont {Ilic}}, \bibinfo {author} {\bibfnamefont {A.}~\bibnamefont
  {Bluestone}}, \bibinfo {author} {\bibfnamefont {N.}~\bibnamefont {Volet}},
  \bibinfo {author} {\bibfnamefont {T.}~\bibnamefont {Komljenovic}}, \bibinfo
  {author} {\bibfnamefont {L.}~\bibnamefont {Chang}}, \bibinfo {author}
  {\bibfnamefont {S.~H.}\ \bibnamefont {Lee}}, \bibinfo {author} {\bibfnamefont
  {D.~Y.}\ \bibnamefont {Oh}}, \bibinfo {author} {\bibfnamefont {M.-G.}\
  \bibnamefont {Suh}}, \bibinfo {author} {\bibfnamefont {K.~Y.}\ \bibnamefont
  {Yang}}, \bibinfo {author} {\bibfnamefont {M.~H.~P.}\ \bibnamefont
  {Pfeiffer}}, \bibinfo {author} {\bibfnamefont {T.~J.}\ \bibnamefont
  {Kippenberg}}, \bibinfo {author} {\bibfnamefont {E.}~\bibnamefont {Norberg}},
  \bibinfo {author} {\bibfnamefont {L.}~\bibnamefont {Theogarajan}}, \bibinfo
  {author} {\bibfnamefont {K.}~\bibnamefont {Vahala}}, \bibinfo {author}
  {\bibfnamefont {N.~R.}\ \bibnamefont {Newbury}}, \bibinfo {author}
  {\bibfnamefont {K.}~\bibnamefont {Srinivasan}}, \bibinfo {author}
  {\bibfnamefont {J.~E.}\ \bibnamefont {Bowers}}, \bibinfo {author}
  {\bibfnamefont {S.~A.}\ \bibnamefont {Diddams}},\ and\ \bibinfo {author}
  {\bibfnamefont {S.~B.}\ \bibnamefont {Papp}},\ }\bibfield  {title} {\bibinfo
  {title} {An optical-frequency synthesizer using integrated photonics},\
  }\href {https://doi.org/10.1038/s41586-018-0065-7} {\bibfield  {journal}
  {\bibinfo  {journal} {Nature}\ }\textbf {\bibinfo {volume} {557}},\ \bibinfo
  {pages} {81} (\bibinfo {year} {2018})}\BibitemShut {NoStop}%
\bibitem [{\citenamefont {Zhu}\ \emph {et~al.}(2021)\citenamefont {Zhu},
  \citenamefont {Shao}, \citenamefont {Yu}, \citenamefont {Cheng},
  \citenamefont {Desiatov}, \citenamefont {Xin}, \citenamefont {Hu},
  \citenamefont {Holzgrafe}, \citenamefont {Ghosh}, \citenamefont
  {Shams-Ansari}, \citenamefont {Puma}, \citenamefont {Sinclair}, \citenamefont
  {Reimer}, \citenamefont {Zhang},\ and\ \citenamefont {Lon\v{c}ar}}]{Zhu:21}%
  \BibitemOpen
  \bibfield  {author} {\bibinfo {author} {\bibfnamefont {D.}~\bibnamefont
  {Zhu}}, \bibinfo {author} {\bibfnamefont {L.}~\bibnamefont {Shao}}, \bibinfo
  {author} {\bibfnamefont {M.}~\bibnamefont {Yu}}, \bibinfo {author}
  {\bibfnamefont {R.}~\bibnamefont {Cheng}}, \bibinfo {author} {\bibfnamefont
  {B.}~\bibnamefont {Desiatov}}, \bibinfo {author} {\bibfnamefont {C.~J.}\
  \bibnamefont {Xin}}, \bibinfo {author} {\bibfnamefont {Y.}~\bibnamefont
  {Hu}}, \bibinfo {author} {\bibfnamefont {J.}~\bibnamefont {Holzgrafe}},
  \bibinfo {author} {\bibfnamefont {S.}~\bibnamefont {Ghosh}}, \bibinfo
  {author} {\bibfnamefont {A.}~\bibnamefont {Shams-Ansari}}, \bibinfo {author}
  {\bibfnamefont {E.}~\bibnamefont {Puma}}, \bibinfo {author} {\bibfnamefont
  {N.}~\bibnamefont {Sinclair}}, \bibinfo {author} {\bibfnamefont
  {C.}~\bibnamefont {Reimer}}, \bibinfo {author} {\bibfnamefont
  {M.}~\bibnamefont {Zhang}},\ and\ \bibinfo {author} {\bibfnamefont
  {M.}~\bibnamefont {Lon\v{c}ar}},\ }\bibfield  {title} {\bibinfo {title}
  {Integrated photonics on thin-film lithium niobate},\ }\href
  {https://doi.org/10.1364/AOP.411024} {\bibfield  {journal} {\bibinfo
  {journal} {Adv. Opt. Photon.}\ }\textbf {\bibinfo {volume} {13}},\ \bibinfo
  {pages} {242} (\bibinfo {year} {2021})}\BibitemShut {NoStop}%
\bibitem [{\citenamefont {Yi}\ \emph {et~al.}(2022)\citenamefont {Yi},
  \citenamefont {Wang}, \citenamefont {Zhou}, \citenamefont {Zhu},
  \citenamefont {Zhang}, \citenamefont {You}, \citenamefont {Zhang},\ and\
  \citenamefont {Ou}}]{Yi2022SiC}%
  \BibitemOpen
  \bibfield  {author} {\bibinfo {author} {\bibfnamefont {A.}~\bibnamefont
  {Yi}}, \bibinfo {author} {\bibfnamefont {C.}~\bibnamefont {Wang}}, \bibinfo
  {author} {\bibfnamefont {L.}~\bibnamefont {Zhou}}, \bibinfo {author}
  {\bibfnamefont {Y.}~\bibnamefont {Zhu}}, \bibinfo {author} {\bibfnamefont
  {S.}~\bibnamefont {Zhang}}, \bibinfo {author} {\bibfnamefont
  {T.}~\bibnamefont {You}}, \bibinfo {author} {\bibfnamefont {J.}~\bibnamefont
  {Zhang}},\ and\ \bibinfo {author} {\bibfnamefont {X.}~\bibnamefont {Ou}},\
  }\bibfield  {title} {\bibinfo {title} {Silicon carbide for integrated
  photonics},\ }\href {https://doi.org/10.1063/5.0079649} {\bibfield  {journal}
  {\bibinfo  {journal} {Applied Physics Reviews}\ }\textbf {\bibinfo {volume}
  {9}},\ \bibinfo {pages} {031302} (\bibinfo {year} {2022})}\BibitemShut
  {NoStop}%
\bibitem [{\citenamefont {Xiong}\ \emph {et~al.}(2012)\citenamefont {Xiong},
  \citenamefont {Pernice}, \citenamefont {Sun}, \citenamefont {Schuck},
  \citenamefont {Fong},\ and\ \citenamefont {Tang}}]{Xiong:12}%
  \BibitemOpen
  \bibfield  {author} {\bibinfo {author} {\bibfnamefont {C.}~\bibnamefont
  {Xiong}}, \bibinfo {author} {\bibfnamefont {W.~H.~P.}\ \bibnamefont
  {Pernice}}, \bibinfo {author} {\bibfnamefont {X.}~\bibnamefont {Sun}},
  \bibinfo {author} {\bibfnamefont {C.}~\bibnamefont {Schuck}}, \bibinfo
  {author} {\bibfnamefont {K.~Y.}\ \bibnamefont {Fong}},\ and\ \bibinfo
  {author} {\bibfnamefont {H.~X.}\ \bibnamefont {Tang}},\ }\bibfield  {title}
  {\bibinfo {title} {Aluminum nitride as a new material for chip-scale
  optomechanics and nonlinear optics},\ }\href
  {https://doi.org/10.1088/1367-2630/14/9/095014} {\bibfield  {journal}
  {\bibinfo  {journal} {New Journal of Physics}\ }\textbf {\bibinfo {volume}
  {14}},\ \bibinfo {pages} {095014} (\bibinfo {year} {2012})}\BibitemShut
  {NoStop}%
\bibitem [{\citenamefont {Shang}\ \emph {et~al.}(2021)\citenamefont {Shang},
  \citenamefont {Wan}, \citenamefont {Selvidge}, \citenamefont {Hughes},
  \citenamefont {Herrick}, \citenamefont {Mukherjee}, \citenamefont {Duan},
  \citenamefont {Grillot}, \citenamefont {Chow},\ and\ \citenamefont
  {Bowers}}]{shang2021perspectives}%
  \BibitemOpen
  \bibfield  {author} {\bibinfo {author} {\bibfnamefont {C.}~\bibnamefont
  {Shang}}, \bibinfo {author} {\bibfnamefont {Y.}~\bibnamefont {Wan}}, \bibinfo
  {author} {\bibfnamefont {J.}~\bibnamefont {Selvidge}}, \bibinfo {author}
  {\bibfnamefont {E.}~\bibnamefont {Hughes}}, \bibinfo {author} {\bibfnamefont
  {R.}~\bibnamefont {Herrick}}, \bibinfo {author} {\bibfnamefont
  {K.}~\bibnamefont {Mukherjee}}, \bibinfo {author} {\bibfnamefont
  {J.}~\bibnamefont {Duan}}, \bibinfo {author} {\bibfnamefont {F.}~\bibnamefont
  {Grillot}}, \bibinfo {author} {\bibfnamefont {W.~W.}\ \bibnamefont {Chow}},\
  and\ \bibinfo {author} {\bibfnamefont {J.~E.}\ \bibnamefont {Bowers}},\
  }\bibfield  {title} {\bibinfo {title} {Perspectives on advances in quantum
  dot lasers and integration with {Si} photonic integrated circuits},\ }\href
  {https://pubs.acs.org/doi/10.1021/acsphotonics.1c00707} {\bibfield  {journal}
  {\bibinfo  {journal} {ACS photonics}\ }\textbf {\bibinfo {volume} {8}},\
  \bibinfo {pages} {2555} (\bibinfo {year} {2021})}\BibitemShut {NoStop}%
\bibitem [{\citenamefont {Ramadan}\ and\ \citenamefont
  {Osgood}(1998)}]{ramadan1998adiabatic}%
  \BibitemOpen
  \bibfield  {author} {\bibinfo {author} {\bibfnamefont {T.~A.}\ \bibnamefont
  {Ramadan}}\ and\ \bibinfo {author} {\bibfnamefont {R.~M.}\ \bibnamefont
  {Osgood}},\ }\bibfield  {title} {\bibinfo {title} {Adiabatic couplers: design
  rules and optimization},\ }\href
  {https://ieeexplore.ieee.org/document/661021} {\bibfield  {journal} {\bibinfo
   {journal} {Journal of lightwave technology}\ }\textbf {\bibinfo {volume}
  {16}},\ \bibinfo {pages} {277} (\bibinfo {year} {1998})}\BibitemShut
  {NoStop}%
\bibitem [{\citenamefont {Jin}\ \emph {et~al.}(2020)\citenamefont {Jin},
  \citenamefont {John}, \citenamefont {Bauters}, \citenamefont {Bosch},
  \citenamefont {Thibeault},\ and\ \citenamefont {Bowers}}]{jin2020deuterated}%
  \BibitemOpen
  \bibfield  {author} {\bibinfo {author} {\bibfnamefont {W.}~\bibnamefont
  {Jin}}, \bibinfo {author} {\bibfnamefont {D.~D.}\ \bibnamefont {John}},
  \bibinfo {author} {\bibfnamefont {J.~F.}\ \bibnamefont {Bauters}}, \bibinfo
  {author} {\bibfnamefont {T.}~\bibnamefont {Bosch}}, \bibinfo {author}
  {\bibfnamefont {B.~J.}\ \bibnamefont {Thibeault}},\ and\ \bibinfo {author}
  {\bibfnamefont {J.~E.}\ \bibnamefont {Bowers}},\ }\bibfield  {title}
  {\bibinfo {title} {Deuterated silicon dioxide for heterogeneous integration
  of ultra-low-loss waveguides},\ }\href
  {https://www.osapublishing.org/ol/fulltext.cfm?uri=ol-45-12-3340&id=432575}
  {\bibfield  {journal} {\bibinfo  {journal} {Optics Letters}\ }\textbf
  {\bibinfo {volume} {45}},\ \bibinfo {pages} {3340} (\bibinfo {year}
  {2020})}\BibitemShut {NoStop}%
\end{thebibliography}%



% \pagebreak

\bigskip
\bigskip

\begin{large}
\noindent \textbf{Methods} 
\end{large}

\medskip

\noindent \textbf{3D mode transitions}

Using a series of adiabatic tapers \cite{ramadan1998adiabatic}, the optical mode is transferred over a vertical distance exceeding 4 \textmu m between an InP/Si hybrid mode in the laser active region, to the ULL SiN waveguide layer of the ultra-high $Q$ resonator. The mode is transferred first from the InP/Si hybrid waveguide into a silicon waveguide and subsequently from the Si waveguide into a SiN redistribution layer (RDL). As these InP, Si, and SiN layers are fabricated either in contact (in the case of InP and Si) or in close proximity (in the case of the Si and SiN RDL), the optical mode transfers rapidly between them and their tapering lengths are less than 300~\textmu m in total. In particular, the InP to Si transition can be very short, as InP and Si feature similar refractive indices. The Si waveguide is subsequently tapered to a very narrow width of less than 200~nm to match the effective index of the RDL SiN waveguide for efficient power transfer. In order to span the vertical distance separating the SiN RDL and the ULL SiN layer on which the high $Q$ resonators reside, the optical mode is gradually evolved from the upper SiN to the lower SiN layer. The RDL SiN and ULL SiN waveguides feature identical core thicknesses of 100~nm, so that their effective indices are readily matched. The RDL SiN waveguide width is thus tapered from 2800 nm to 200 nm, while simultaneously widening the ULL SiN waveguide width from 200 nm to 2800 nm, over a distance approaching 1~cm in length. This scheme enables efficient power transfer ($<$ 1 dB insertion loss) from the RDL SiN waveguide to the ULL SiN waveguide.

\medskip
\noindent \textbf{Device fabrication}
The device fabrication starts with a 200-mm-diameter Si wafer with 15-\textmu m thick thermal silicon dioxide (SiO$_2$). Low-pressure chemical vapor deposition (LPCVD) silicon nitride with 100-nm thickness is deposited and serves as the ULL layer before SiN waveguide definition. Multiple rounds of tetraethyl orthosilicate (TEOS)-based oxide are deposited on the ULL layer to form an approximately 4 \textmu m thick spacer layer, followed by another 100~nm-thick LPCVD SiN deposition. The adiabatic RDL taper is defined and etched on this layer. TEOS oxide is deposited again with chemical mechanical polishing (CMP) process to leave around 500 nm thick SiO$_2$ on top of RDL. The wafer is cored into 100-mm wafers to be compatible with an ASML 248-nm DUV stepper. Diced silicon-on-insulator (SOI) pieces are bonded on the polished SiO$_2$ surface using plasma-activated direct bonding. The Si substrate is removed by mechanical polishing plus deep Si Bosch etching. The buried SiO$_2$ (BOX) layer is removed by buffered hydrofluoric acid (BHF). The fabricated Si/SiN RDL/SiN ULL wafer is then ready for the heterogeneous InP process on Si similar to our previous works~\cite{Xiang:20}. Generally, Si waveguides and tapers are patterned with DUV stepper while the grating is patterned with electron beam lithography with a period of 240 nm. InP dies with layer stack shown in Fig.~\ref{Fig:1}c are bonded on the fabricated Si circuits, with the InP substrate removed by mechanical polishing and 3:1 Hydrochloric acid: deionized water (HCl:DI). A thin layer of P-type contact metal Pd/Ge/Pd/Au is formed using liftoff process. The InP mesa is etched using CH$_4$/H$_2$/Ar, with an SiO$_2$ hard mask. The dry etch is monitored using an etch monitor and stops at the AlInGaAs QW layer. After another round of QW layer lithography, the QW layer is etched using a mixed solution of H$_2$O/H$_2$O$_2$/H$_3$PO$_4$ 15/5/1. N-type InP mesa etch follows QW etch to complete the mesa definition. The excess Si on top of the SiN devices is removed using XeF$_2$. The entire chip is passivated using low-temperature deuterated SiO$_2$~\cite{jin2020deuterated} followed by the contact metal window opening. N-type contact metal Pd/Ge/Pd/Au and another layer of Ti/Au on top of the P-type contact metal are deposited and formed. The chip goes through another round of SiO$_2$ deposition and contact via opening. Proton implantation is performed to define the current channels. Ti/Pt is deposited as heaters for the phase tuner on Si and resonance tuner on SiN. Ti/Au probe metal is deposited to finish the wafer fabrication. The fabricated 100-mm-diameter 3D PIC wafer is diced and polished to expose the SiN edge couplers for fiber-coupled device characterization. 

\medskip
\noindent \textbf{Laser self-injection locking}
By tuning the laser wavelength to a resonance from the ring, the back-scattered light from the ring locks the laser wavelength to the resonance provided that the phase of the back-scattered light arriving at the laser is an integer multiple of 2 $\pi$ of the forward laser output phase. In other words, the wavelengths of the laser and the resonance are matched in the frequency domain while the phases of the laser and the back-scattered light are matched in the time domain as shown in Fig.~\ref{Fig:2}a. Matching the wavelengths is performed by tuning either the laser gain current or the ring heater current, while matching the phases is done by tuning the phase tuner current. Both laser gain current and phase tuner current are driven with low-noise laser current sources (ILX Lightwave LDX-3620) to ensure stable and low-noise operation. Detection of the self-injection locking state is assured by not only observing the decrease in the output power from the ring when the laser wavelength hits the resonance but also the decrease in the linewidth of the self-heterodyning beat as the self-injection locking takes place as shown in Fig.~\ref{Fig:2}b. The self-heterodyne interferometer setup consists basically of a Mach-Zehnder interferometer (made from two 3 dB couplers) with a polarization controller and a short delay line in one of its arms and a fiber-coupled acoustic-optic modulator (Gooch \& Housego 27 MHz) in the other arm, as shown in Fig.~\ref{Fig:2}b. The beat frequency from the self-heterodyne interferometer is detected using a photodetector (Newport 1811) before it is sent to an Electrical Spectrum Analyzer (ESA) (Rohde \& Schwarz FSWP). 
The phase tuner current is roughly adjusted during the self-injection locking process to allow the locking to occur and finely tuned afterwards to ensure stable self-injection locking. It is worth mentioning here that the self-injection locking state can last for hours without even packaging the laser chip. This can be attributed to the integration of the laser and the resonator on the same chip which reduces the phase fluctuation between the laser and the back-scattered light from the ring. 

\medskip
\noindent \textbf{SIL laser characterization}
The dynamics of the phase tuner influence on SIL is investigated by sweeping its applied electrical power (Keithley 2604B) over three 2 $\pi$ injection locking periods, while recording the ESA spectrogram of the detected self-heterodyning beat of the SIL laser (Fig.~\ref{Fig:2}c top). The spectrogram of the injection-locking periods, which is depicted in Fig.~\ref{Fig:2}c, demonstrates stable SIL periods (dark blue regions) followed by chaotic regions (light blue) and then unlocked regions. The laser power is detected also on an oscilloscope (Tektronix MSO64B) during the phase tuning over only one period, which clearly shows the mentioned behavior (Fig.~\ref{Fig:2}c bottom). 
Another important parameter is the frequency range at which the self-injection locking persists. It can be obtained by either sweeping the laser frequency over the ring resonance or sweeping the ring resonance over the laser. We selected the second scheme by sweeping the current of the ring heater using a triangle signal applied to the current source (Keithley 2604B). To detect the change in laser linewidth during sweeping, a beat is made using 3 dB coupler between the SIL laser and a narrow-linewidth fiber laser. The beat is optically amplified with an Erbium-Doped Fiber Amplifier (EDFA) (Amonics AEDFA-IL-18-B-FA) and sent to the fast photodetector (Finisar HPDV2120R) that is connected to the ESA, as shown in the lower branch of Fig.~\ref{Fig:2}b. The recorded spectrogram during the resonance sweep is shown in Fig.~\ref{Fig:2}d.

\medskip
\noindent \textbf{Laser feedback sensitivity measurement}
Figure \ref{Fig:3}b schematically depicts the experimental configurations for analyzing the feedback sensitivity of laser. The coupled laser emission is sent to a 90/10 fiber beam splitter, after which 90\% of the coupled power will be used for external optical feedback. The feedback loop consists of an 8-meter-long single-mode fiber (SMF), a three-port optical circulator (CIR), a polarization controller (PC), and a variable optical attenuator (VOA) that allows for an attenuation ranging from 0 to 40 dB (EXFO MOA-3800). It should be noted that the laser feedback sensitivity also depends on the polarization of the reflected field, which must be adjusted to maximize the feedback influence before running the analysis. The remaining 10\% of laser output is utilized for the feedback sensitivity characterization. After passing through an optical isolator, it is transferred either to a phase noise analyzer (PNA) (OEWaves OE4000) for frequency noise characterization or a delayed self-heterodyne setup for the electrical spectrum-based laser coherence check.

In this study, the feedback strength is determined by the reflected power ($P_{refl}$) and the free-space output power ($P_{out}$) through the following relationship: 
\begin{equation}
\eta_{F}=\frac{P_{refl}}{P_{out}}
\label{equ:fb}
\end{equation}

All losses from the feedback loop should be considered to calculate the reflected power, thus the feedback strength. After optimizing the setup, the fiber-chip coupling loss is -3 dB (round-trip coupling loss is -6 dB), the total losses from the 90\% beam splitter, the optical circulator, the insertion loss of the VOA, the polarization controller, and the fiber is -4.05 dB. The feedback strength that accounts for the attenuation of VOA can thus be tuned from -10.05 to -50.05 dB.

To further reduce the loss from setup and thus to maximize the feedback strength as large as possible, we use a 100\% fiber backreflector (BKR, Thorlabs) to replace the configurations after the 90\% beam splitter port. The loss from the feedback loop is then reduced to -0.9 dB, and the maximum feedback strength is -6.9 dB.

\medskip
\noindent \textbf{Microwave signal generation}
Two lasers are self-injection locked to two 30-GHz-FSR ring resonators that are shifted in frequency by 10 GHz without tuning on the ring resonance. Although each ring resonator can be tuned over 30 GHz by applying around 0.5 W of electrical power to the heater on the ring, we used only one ring heater for tuning. The tuning range for the first FSR is -10 GHz to 20 GHz, while the next full-FSR tuning leads to 20 GHz to 50 GHz tuning by locking the second laser to the next resonance and hence covering the full 50 GHz. As demonstrated in Fig.~\ref{Fig:4}b. The two lasers’ outputs are collected from the chip by the VGA and sent to a 3-dB fiber coupler, then EDFA before beating on a fast PD (Finisar HPDV2120R) connected to ESA. Small portion of the lasers (1\%) is sent to an Optical Spectrum Analyzer (OSA) (Yokogawa AQ6370C) for monitoring if the lasers are on the intended resonance. Although our chip could be used to generate any arbitrary microwave frequency over the 50 GHz, it is used here to generate microwave frequencies at steps of 1 GHz over the full 50 GHz for demonstration (Fig.~\ref{Fig:4}c). An offset phase locking servo circuit (Vescent D2-135) is used to improve the long-term stability of the generated microwave signals for frequencies up to 10 GHz, by locking the phase of one of the lasers to the other one. The feedback signal from the servo control box is sent to one of the laser’s ring heaters to lock its phase to the phase of the second laser. Stable microwave signals are thus generated with low phase noise characteristics of the SIL lasers.   

\end{document}